\documentclass{article}
\usepackage{epsfig}
\usepackage{english}
\usepackage{wrapfig}
\usepackage{afterpage}
\renewcommand{\floatpagefraction}{4}
\renewcommand{\textfraction}{0.1}
\renewcommand{\topfraction}{0.9}
\renewcommand{\bottomfraction}{0.9}
\begin{document}

\renewcommand{\floatpagefraction}{4}
\renewcommand{\textfraction}{0.1}
\renewcommand{\topfraction}{0.8}
\renewcommand{\bottomfraction}{0.8}

\begin{center}
   {\large \bf Dynamics of Multifragmentation in Heavy Ion Collisions} \\[5mm]
   Willibrord.~REISDORF\footnote{ E-mail address: W.Reisdorf@gsi.de} \\[5mm]
   {\small \it  Gesellschaft f\"ur Schwerionenforschung (GSI) \\
   Postfach 110552, D-64220 Darmstadt, Germany \\[8mm] }
\end{center}

\begin{abstract}\noindent
We review multifragmentation data obtained at the SIS/GSI accelerator using
heavy ion beams with (0.1-1)A GeV together with the ALADIN and FOPI
experimental setups.
\end{abstract}
\large
\section{Introduction}
Multifragmentation, that is the emission of several intermediate mass
fragments from a hot nucleus, is a phenomenon observed in  nuclear
reactions, using light and heavy projectiles over a wide incident enery
range.
Besides the scientific curiosity to understand what happens, some of the
hopes in these studies are that one can learn more about the tendency of
fermionic nuclear matter to appear in clusters and, perhaps eventually,
about the topology of the nuclear phase diagram, in particular the evasive
liquid-to-gas transition and, from the explosive features of some of the
reactions, about nuclear matter compressibility.
Such information is of high interest in other fields as well, such as
astrophysics.
As the underlying mechanisms are likely to be rather complex, statistical
approaches have been invoked from the beginning in order to try to
understand the experimental data.
Evidence for non-equilibrium effects have lead to the parallel development
of sophisticated transport models as well.

As this written contribution represents essentially the contents of a
one-hour lecture, it is highly limited in scope:  I 
describe experimental multifragmentation data measured in the past years at
GSI, Darmstadt (Germany).
The many facets of multifragmentation are studied in laboratories all over
the world, however.
I refer the reader to a recent workshop~\cite{Hirschegg99} on
Multifragmentation for more encompassing information on this subject.

I shall start by describing the ALADIN and FOPI
experimental setups and the way observed
multiparticle events are sorted. 
Then I review fragment yield distributions in both central and peripheral
collisions together with attempts to derive from them apparent chemical
temperatures.
Detailed information on momentum space distributions will allow me to
discuss subjects such as the determination of the caloric curve, the
influence of flow and the system-size dependence.
Finally, an isospin tracer experiment, performed to study the degree of
equilibration, will be described.

\section{Apparatus}
\begin{figure}[!h]
\begin{center}
\epsfig{file=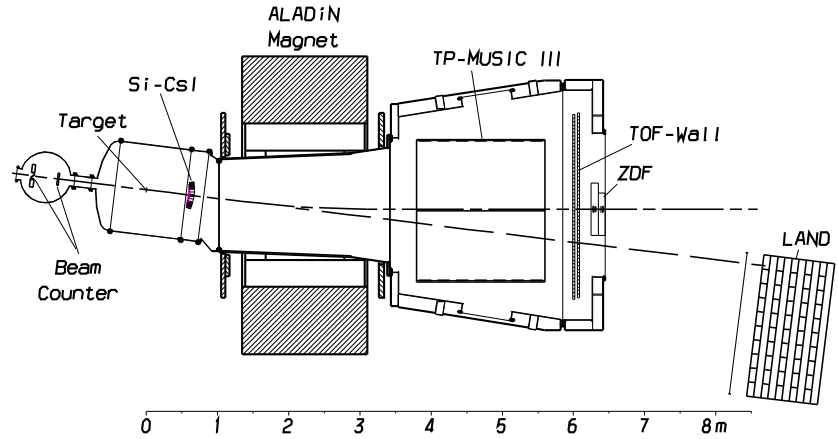,width=11cm}

\caption{
Schematic view of the ALADIN spectrometer in the bending 
plane~\protect\cite{Schuettauf96}.
}
\label{ALADIN}
\end{center}
\end{figure}

\begin{figure}[!h]
\begin{center}
\epsfig{file=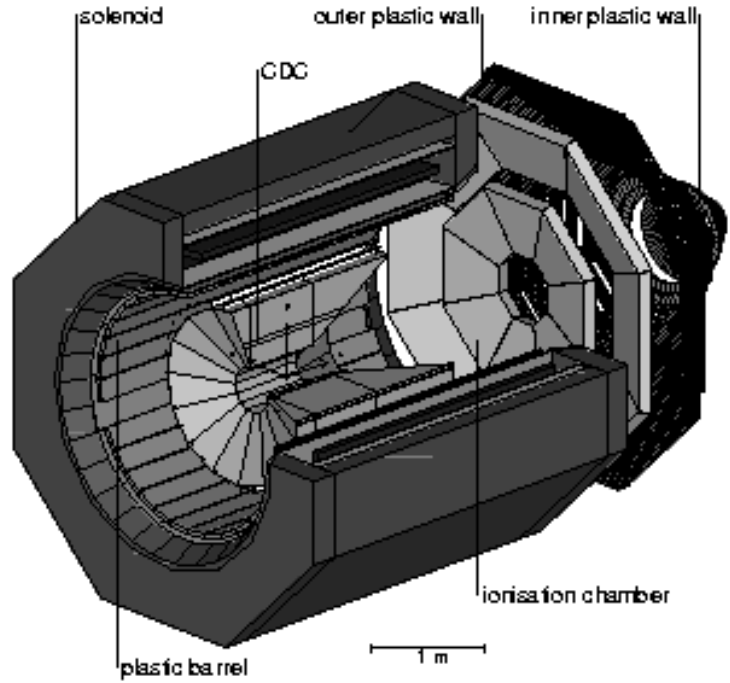,width=12cm}

\vspace{-1.5cm}
\caption{
Schematic view of the FOPI detector.
}
\label{FOPI}
\end{center}
\end{figure}

Two large detection systems have been set up at the heavy ion synchrotron
SIS at GSI Darmstadt, ALADIN (Fig.~\ref{ALADIN})
 and FOPI (Fig.~\ref{FOPI}).              

\vspace{-0cm}

\begin{figure}[!h]
\parbox{6.6cm}{
\epsfig{file=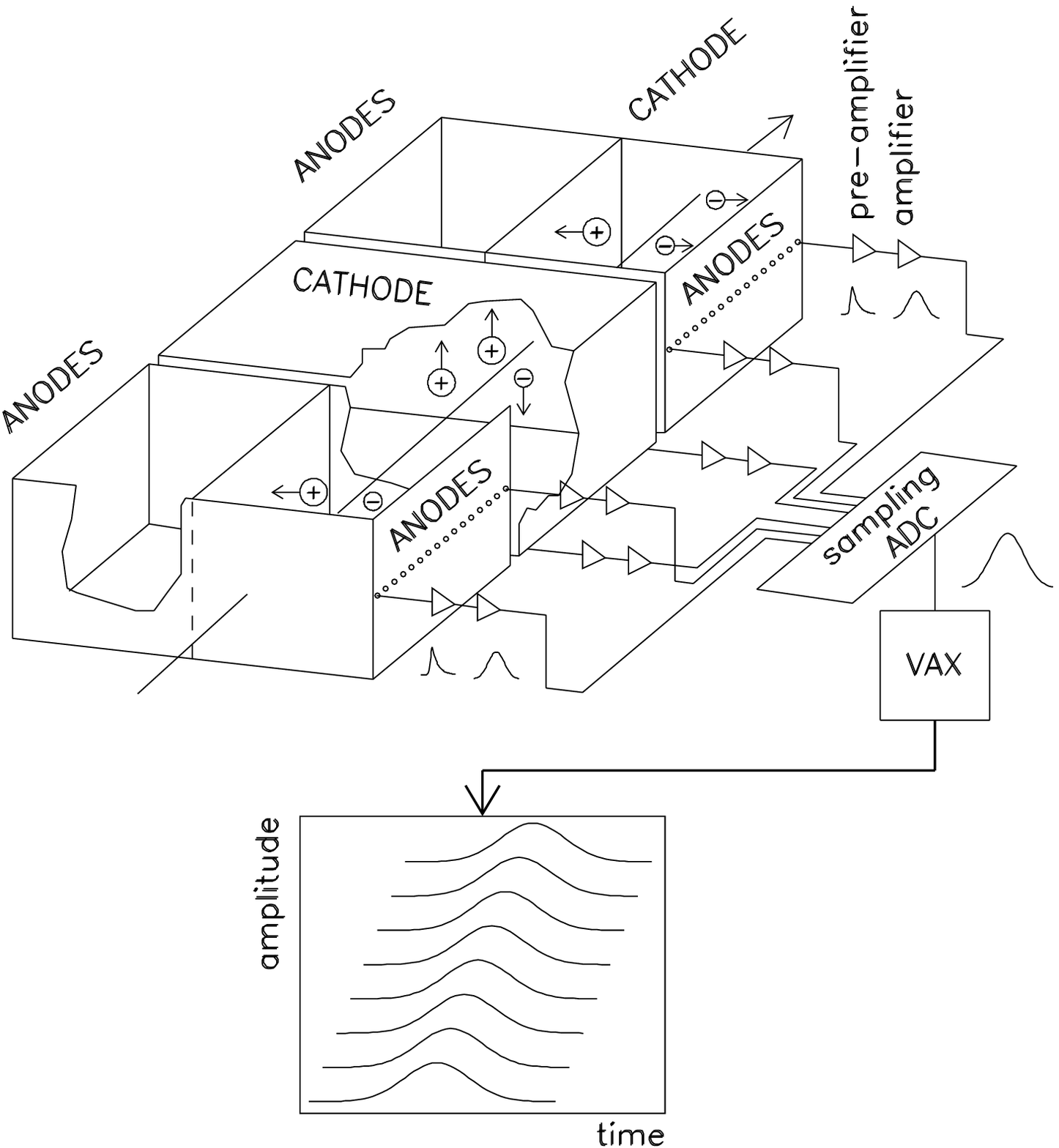,width=6.6cm}

\caption{
Illustration of the operation of the MUSIC II detector
From ref.~\protect\cite{Begemann98} 
}
\label{MUSIC1}
}
\hspace{8mm}
\parbox{6.6cm}{
\epsfig{file=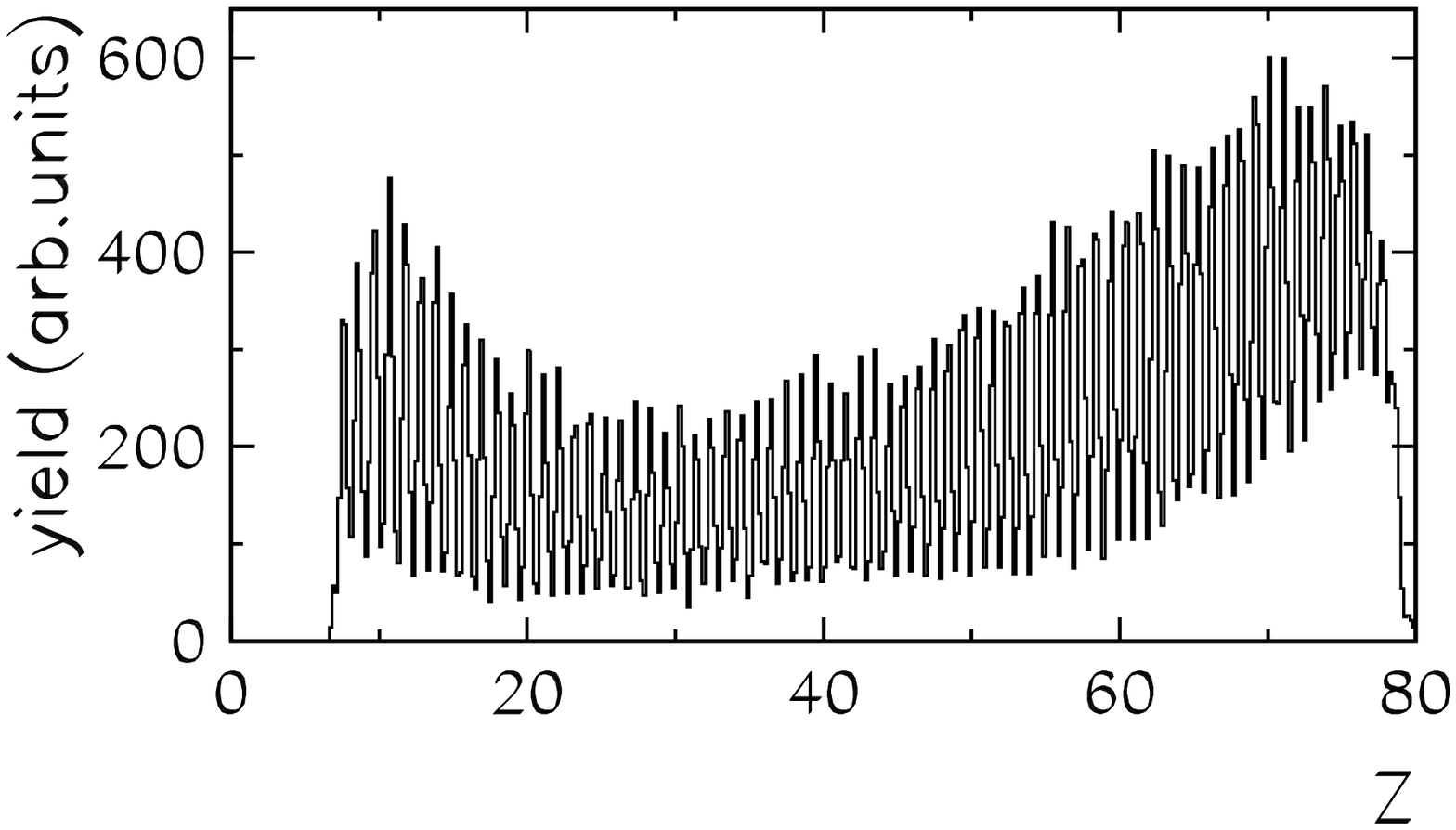,width=6.6cm}

\caption{      
Observed charge spectrum.
From ref.~\protect\cite{Begemann98}  
}
\label{MUSIC2}
}
\end{figure}

The ALADIN~\cite{Schuettauf96} spectrometer was designed primarily to look
at fast projectile-like spectator fragments focussed into a forward cone of
about $\pm 5^\circ$. The core of the apparatus consists of a magnet
operated at a bending power of 1.4 Tm, a time projection multiple-sampling
ionization chamber TP-MUSIC and a time-of-flight (TOF) wall.
In some experiments complementary detectors are installed, such
as a large area neutron detector LAND that identifies neutrons emitted
close to $0^\circ$, and a series of multidetector hodoscopes consisting of
Si-Cs(Tl) telescopes, CaF$_2$ plastic phoswich detectors and Si-strip
detectors arranged so as to cover angles between $6-58^\circ$.
One of the purposes of the TP-MUSIC detector is to allow precise tracking
of charged particles passing through its active volume.
The version MUSIC-II~\cite{Begemann98} is shown in Fig.~\ref{MUSIC1}.                  

 
\vspace{0.2cm}

\vspace{0.4cm}
It consists of
three active volumes with the electric drift fields in adjacent sections
perpendicular to each other, two for the measurement of the horizontal and
one for that of the vertical position and angle of the particle track.
To allow multiple sampling of the particle signals, each anode is
subdivided into 16 stripes with a width of 3 cm each.
The anode signals are recorded using flash analog-to-digital converters
(ADCs) with a sampling rate of 16 MHz.
Better than unit resolution for nuclear charges from Z=8 all the way to
Au (Z=79) is obtained (Fig.~\ref{MUSIC2}).
Charges between Z=2 and 15
can be resolved with use of the scintillator strips in the TOF.

\begin{figure}[!h]
\begin{center}
\epsfig{file=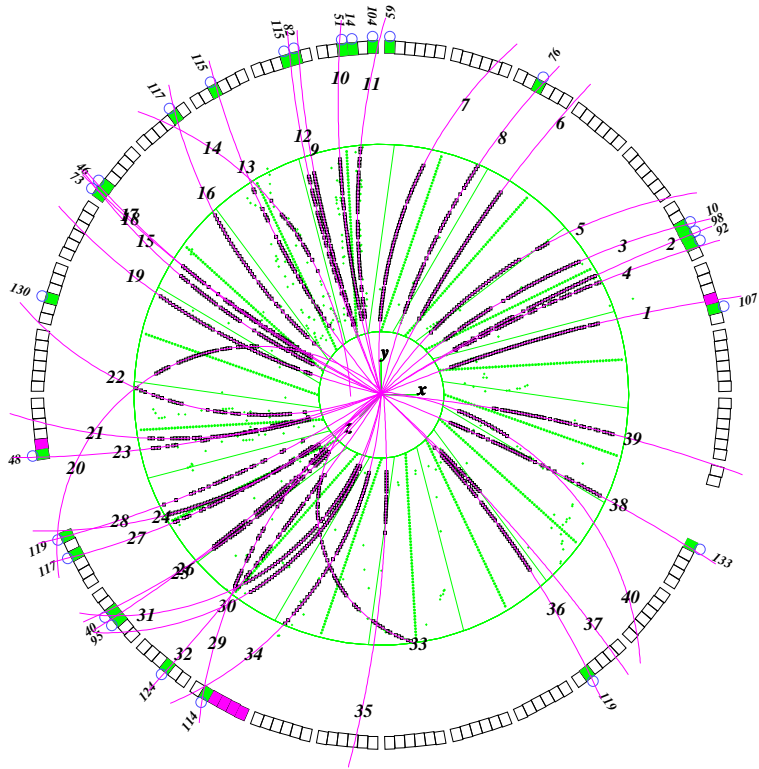,width=10cm}

\vspace{2mm}
\caption{
Event display in the Central Drift Chamber. A cross section in the
plane perpendicular to the beam axis is shown
}
\label{CDCtrack}
\end{center}
\end{figure}

The FOPI apparatus was built to study central heavy ion reactions
in the energy range 0.1-2A GeV.
Particle identification, again, is done by a combined energy-loss,
time-of-flight and magnetic rigidity analysis. The magnet is a
superconducting solenoid operated at 0.6 T (see Fig.~\ref{FOPI}).
Time of flight and energy-loss are determined by about 1000 scintillator
detectors arranged octogonally in the downstream part of the detector
(PLASTIC WALL) and as a barrel inside the magnet.
At lower incident energies (E/A $\leq 400$ MeV) a set of gas ionization
chambers is inserted in front of the PLASTIC WALL to allow identification
of slower heavy clusters (up to Z about 15).
Inside the magnet, the particles are tracked in drift chambers,
the Central Drift Chamber, CDC, and the HELITRON (usually installed in
front of the CDC, but not shown here).
Fig.~\protect\ref{CDCtrack} 
allows to visualize individual
particle tracks in a specific event determined with use of local or global
(Hough transform) track finding methods.

\begin{figure}[!t] 
\parbox{6.6cm}    {
\hspace{-1cm}
\epsfig{file=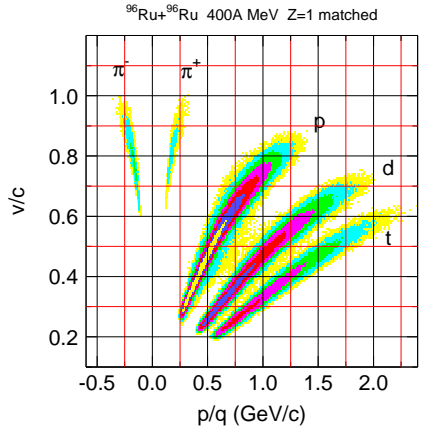,width=6.6cm}
}
\hspace{8mm}
\parbox{6.6cm}{
\caption{
Identification of charge-one particles in the velocity versus 
magnetic rigidity plane
}
\label{CDCpid}
}
\end{figure}

By matching these tracks to the outer barrel scintillators one obtains
the energy loss (from the ionization in the CDC gas), the
track curvature and the time-of-flight (from the Barrel ).
Fig.~\protect\ref{CDCpid}                  
shows well separated pions ,
protons, deuterons and tritons. 


 \newpage
\section{Event sorting}
\begin{wrapfigure}{l}{6.6cm}     
\epsfig{file=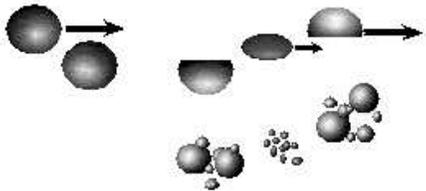,width=3cm,angle=-90}
\caption{Evolution of a heavy ion collision: spectators and participants}
\label{HIC}

\end{wrapfigure}
\begin{figure}[!h]
\parbox{6.6cm}{
\epsfig{file=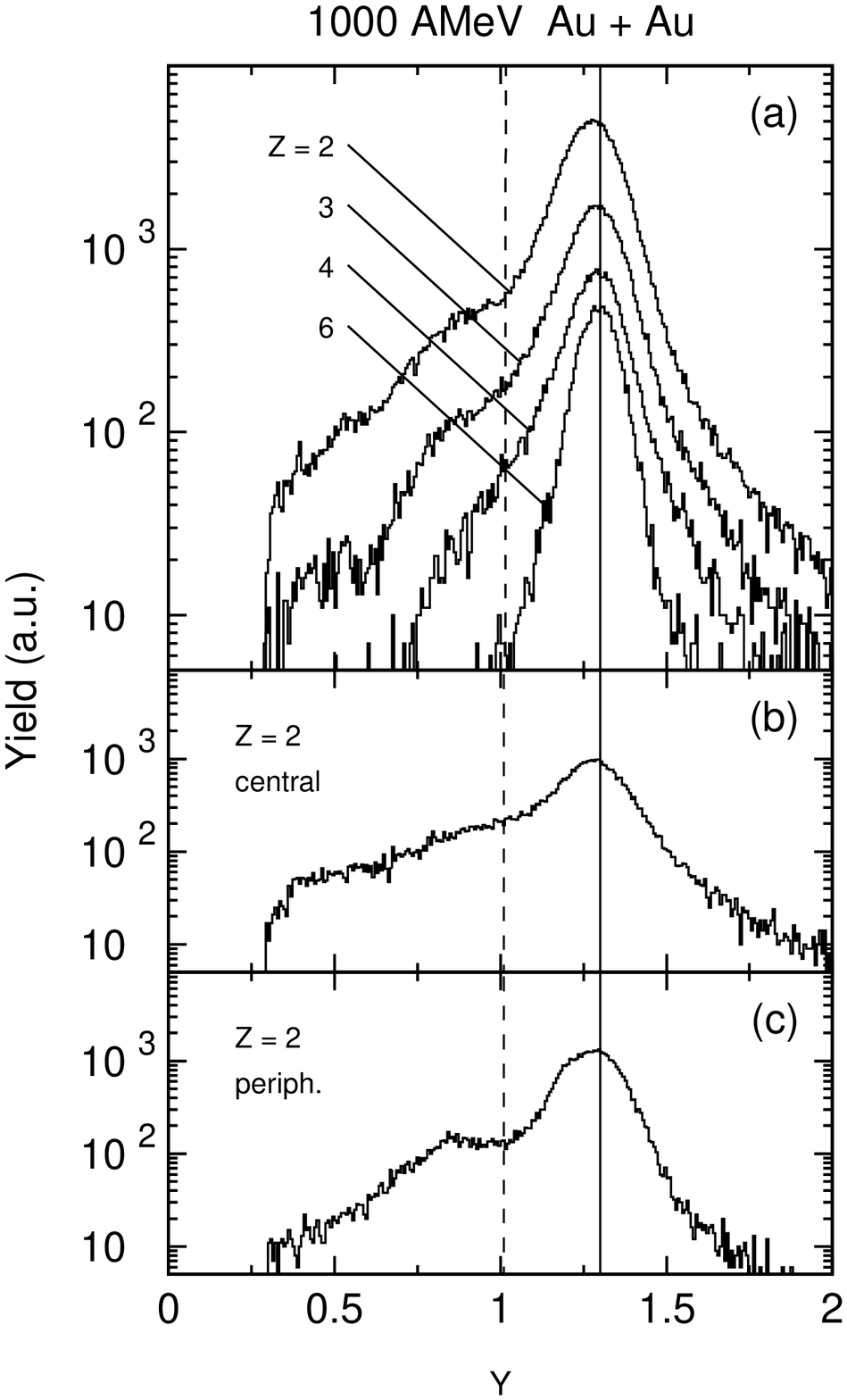,width=6.1cm}
\caption{
(a): Rapidity spectra measured in the reaction Au+Au at
1000A MeV for fragments with Z=2,3,4 and 6. (b): Rapidity spectra for Z=2
fragments in central collisions and (c): in peripheral collisions.
From ref.~\protect\cite{Schuettauf96} 
}
\label{rapidity}
}
\hspace{8mm}
\parbox{6.6cm}{
\vspace{-1.3cm}
\hspace{-0.5cm}
\epsfig{file=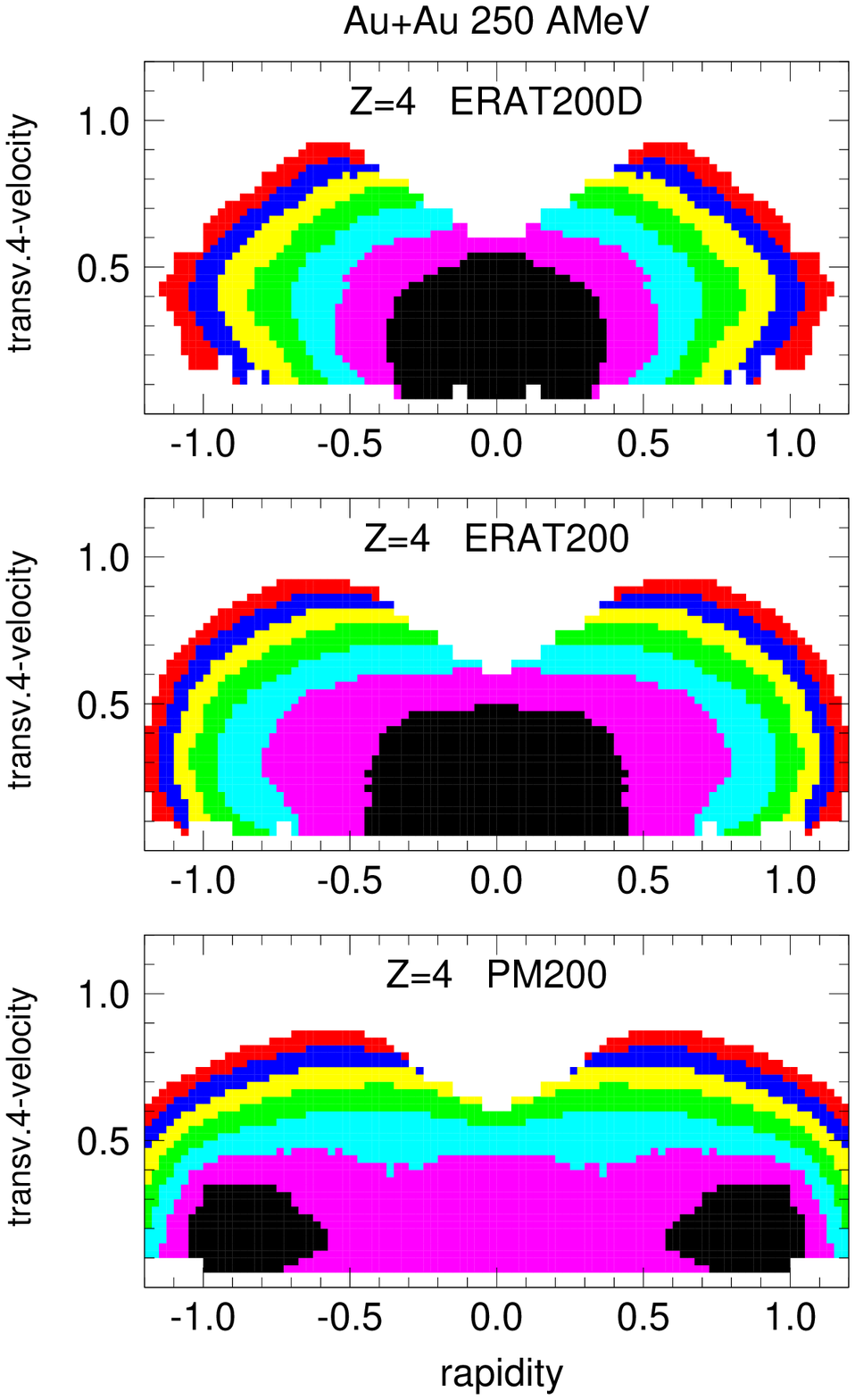,width=8cm}
\caption{
Invariant cross section contour plots (scaled units) for
Be (Z=4) fragments observed in the system Au+Au at 250A MeV. The
three panels correspond to different event selections,
 see text. From ref.~\protect\cite{Reisdorf97}  
}
\label{ptyc}     
}
\end{figure}

Heavy ion collisions lead to events with very different topologies.
It is highly desirable to sort these events according to some criteria,
such as the impact parameter $b$. As $b$ itself cannot be measured directly
one tries to find an observable that is believed to be strongly correlated
to $b$.
The methods used by both ALADIN and FOPI are easiest to visualize in 
the participant-spectator picture illustrated in Fig,~\ref{HIC}.
In experiments with the ALADIN spectrometer one selects impact parameters by
measuring the size of the projectile spectator: one adds up the charges of
all the fragments with $Z >  1$ moving forwards in a narrow rapidity
window, see Fig.~\ref{rapidity}.                   
This sum has been dubbed $Z_{bound}$: large $Z_{bound}$ obviously
correspond to rather peripheral collisions.
In experiments with the FOPI detector which are primarily performed to
study fireball (participant) physics, a standard method is to determine the
charged-particle multiplicity. The underlying idea is that 'fireball
matter' emits more particles than 'spectator matter'.
In the early phase of running the detector the multiplicity, PM,
observed in the PLASTIC WALL, which covers only laboratory angles forwards
of $30^\circ$, was taken.
Due to this geometric limitation, overcome later when the full detector
became available, other more powerful methods to 'maximize' the fireball
(i.e. choosing very central collisions) were developped.  
Fig.~\ref{ptyc} 
illustrates the increasingly 'compact' momentum
space topology of Z=4 fragments emitted in Au on Au reactions at 250A MeV
as selection is done, first, according to high Plastic Wall multiplicities
PM (lower panel, where still spectator peaks are visible), second,
according to large 'ERAT', the ratio of total transverse to longitudinal
kinetic energies (center panel) and \, finally, using large ERAT and low
directivity, i.e. high azimuthal symmetry. 
The directivity is obtained from
$\sum{\omega Z \vec{u_t}}/ \sum{Z |u_t|}$ ($\omega = \pm 1$
forward/backward,
Z fragment charge, $\vec{u_t}$ transverse 4-velocity).
In the figure both the rapidity (abscissa) and the transverse 4-velocity
(ordinate) are scaled with the respective projectile values, placing the
midrapidity value at zero.
The ERAT and PM selections correspond to cross sections of 200 mb, the
additional low-directivity condition cuts the cross section down to 70 mb.
For more details see~\cite{Reisdorf97}.

\afterpage{\clearpage}
\section{Cluster production: peripheral and central collisions.
Universality.}
\begin{figure}[!h]
\parbox{6.0cm}{
\vspace{-1.3cm} 
\hspace{-0.8cm}
\epsfig{file=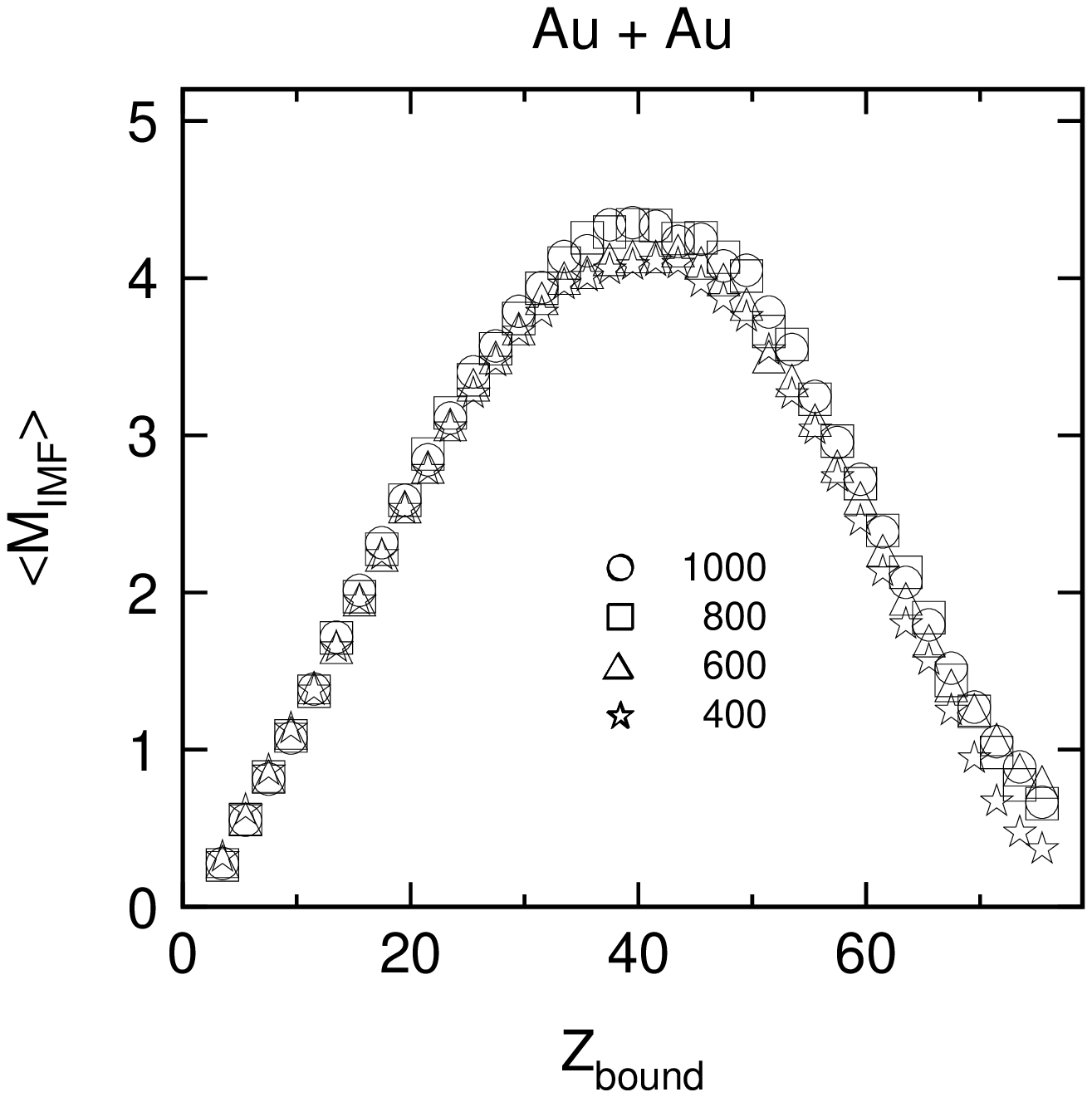,width=6.6cm}
\caption{(a): Mean multiplicity of intermediate mass fragments  
as a function of $Z_{bound}$ for the reaction Au+Au at E/A =
400, 600, 800 and 1000 MeV.~\protect\cite{Schuettauf96}
}
\label{universal}
}
\hspace{6.6mm}
\parbox{7.2cm}{
\hspace{-0.6cm}
\epsfig{file=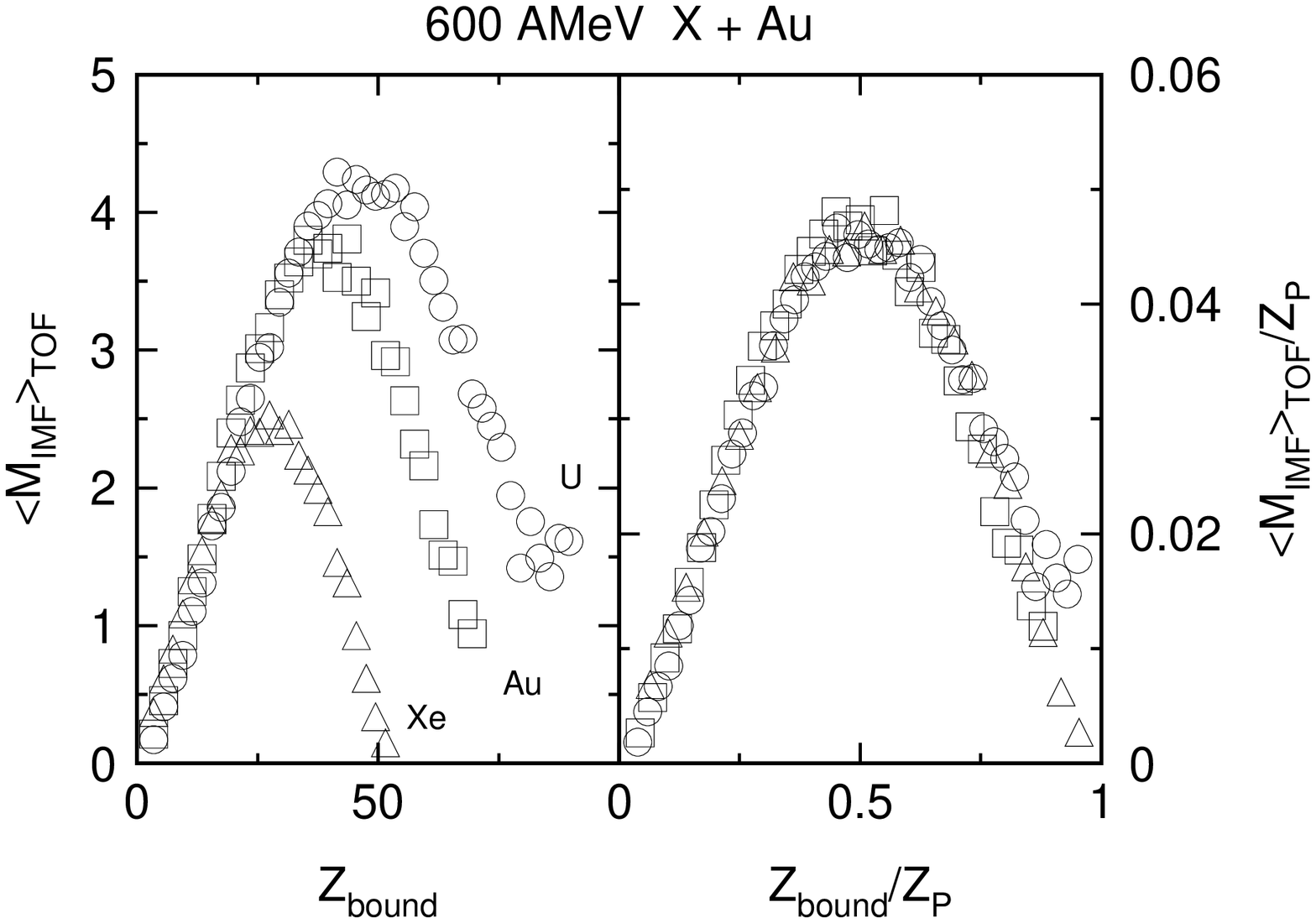,width=8cm}
\vspace{-2mm}
\caption{Left panel: Mean multiplicity of intermediate mass fragments as a
function of $Z_{bound}$ for the reactions $^{238}$U on $^{197}$Au (circles),
$^{197}$Au on $^{197}$Au (squares) and $^{129}$Xe on $^{197}$Au (triangles)
at E/A= 600 MeV.
Right panel: the same data after normalizing both quantities with respect
to the atomic number $Z_p$ of the projectile.~\protect\cite{Schuettauf96}}
\label{universal1}
}
\end{figure}

Multifragment decays of projectile-spectator sources have been studied
extensively with the ALADIN spectrometer.~\cite{Kreutz93}~-~\cite{Odeh99} \
One of the hopes was that spectator matter would be 'gently' excited,
initially uncompressed nuclear matter, that would equilibrate relatively
fast, then expand slowly under heat pressure and enter into the unstable
spinodal regime where its decay would show features of a liquid-to-vapour
transition.
Evidence that multifragmentation was a phenomenon following complete
equilibration was obtained by the observed 'universality' of the decay
pattern.
As shown in Fig.~\ref{universal}, if one plots the observed mean
multiplicity of intermediate mass fragments (IMF, $Z=3-30$) against the
size of the spectator charge , characterized by $Z_{bound}$, (i.e. against
the impact parameter for a fixed target-projectile system), one finds a
dependence that remains invariant against the variation of the incident
energy.~\cite{Schuettauf96} \

Other observables characterizing the population of partition space were
found to be 'universal' as well. 
Provided the projectile beam (Au) was not changed, universality was not
lost when the Au target was replaced by lighter target materials, all the
way down to Be.
Further 'ALADIN plots', obtained by fixing the target to Au, but varying
this time the size of the projectile from U down to Xe (left panel
Fig.~\ref{universal1}),                      
are again merged into one curve when scaling the
IMF multiplicity with the projectile size (right panel
Fig.~\ref{universal1}). 

\begin{figure}[!h]
\begin{center}
\epsfig{file=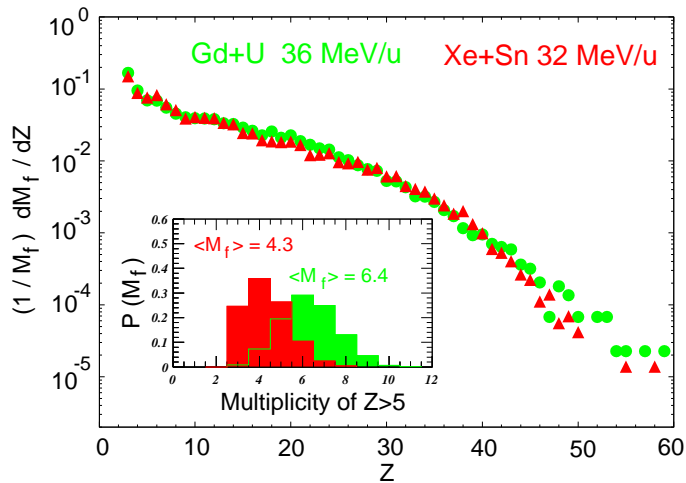,width=9cm}
\end{center}
\caption{Charge distributions in the reactions Gd+U at 36A MeV (circles)
and Xe+Sn at 32A MeV (triangles). Insert: Multiplicity distributions
for fragments with $Z \geq 5$ in the same        
systems. After ref.~\protect\cite{Rivet98}}
\label{Rivet}
\end{figure}

The independence of fragment charge distributions on the size of the source
could also be observed in central fusion-like reactions at lower
incident energies using the $4\pi $ multidetector INDRA at the GANIL
accelerator, Caen, France, see Fig.~\ref{Rivet}.~\cite{Rivet98} \
Studying the systems $^{129}$Xe+Sn at 32A MeV and $^{238}$U+$^{155}$Gd at
36A MeV, it was found that the charge distributions for fragments with
charge $Z > 4$ were essentially identical, if scaled with the system size,
or, as done in the figure, with the average multiplicity of charged
particles with $Z > 4$ (the ratio of the latter is 
close to the ratio of the system sizes).
It was argued in ref.~\cite{Rivet98} that the available excitation
energy, 7A MeV, was the same for both systems.

\begin{figure}{!t}
\parbox{6.6cm}{
\hspace{-1cm}
\epsfig{file=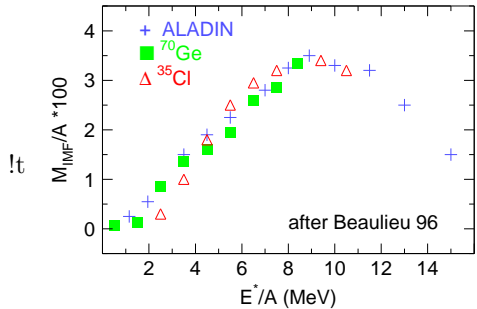,width=7cm}
}
\hspace{8mm}
\parbox{6.6cm}{
\caption{Average number of intermediate mass fragments (scaled by the    
system size) as a function of the excitation energy per          
nucleon. 
Triangles: $^{35}$Cl + $^{197}$Au at 43A MeV, squares: $^{70}$Ge+Ti at 35A
MeV, crosses: ALADIN data. After ref.~\protect\cite{Beaulieu96}} 
\label{Beaulieu}
}
\end{figure}

Beaulieu et al.~\cite{Beaulieu96} pointed out earlier that the
excitation energy per nucleon was the relevant parameter for sorting out
multifragmentation data.
In Fig.~\ref{Beaulieu} the authors show evidence for scaling of
system-size normalized IMF multiplicity data, when plotting them against
the excitation energy per nucleon:
quasi-projectile decay data for the reactions $^{35}$Cl+$^{197}$Au at 43A
MeV and $^{70}$Ge+Ti at 35A MeV are shown together with ALADIN spectator
decay data in the reaction Au+Au at 600A MeV.
The non-trivial determination of the excitation energy deposited into the
presumably thermalized part of the system will be discussed in the next
section.
As pointed out in ref.~\cite{Beaulieu96} there is an important
difference between 'quasi-projectile' data (E/A $< 50$ MeV) and
'projectile-spectatator' data (E/A $\geq  600$ MeV):
in the former case the average mass of the emitting sources is almost
constant due to total-charge requirements in the experiment, whereas in the
latter case the Au projectile spectator decreases from mass $A \sim 190$ to
$A \sim 50$ over the excitation energy range in the figure.



\begin{wrapfigure}{l}{6.6cm}    
\vspace{-1cm}
\hspace{-1cm}
\epsfig{file=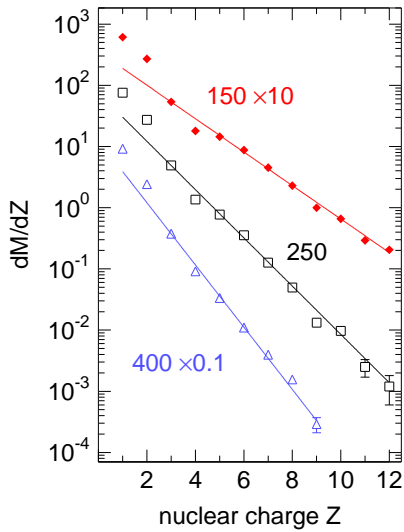,width=6.5cm}
\caption{Charge distributions in central collisions of Au on Au
at 150, 250 and 400A MeV.~\protect\cite{Reisdorf97}}
\label{zdist}
\end{wrapfigure}

\begin{figure}
\begin{center}
\epsfig{file=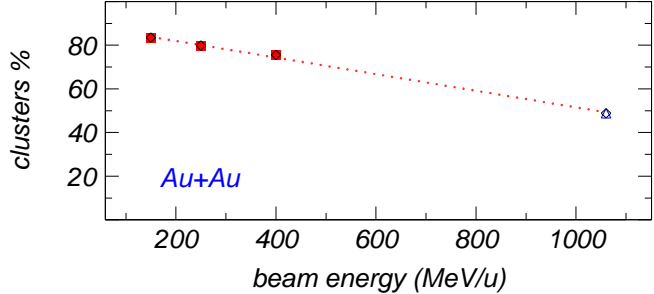,width=10cm}
\end{center}
\caption{Degree of clusterization $(\%)$ of protons in central Au on Au
collisions.}
\label{cluster}
\end{figure}

Before closing this section, let us briefly present the 'fireball' part of
the story.~\cite{Reisdorf97} \
In very central Au on Au collisions above 100A MeV one finds almost
perfectly exponential charge distributions (Fig.~\ref{zdist}), the slope
merely becoming steeper as the excitation energy is raised.
As shown in the Fig.~\ref{cluster}, the probability for a proton to
appear attached to at least one other nucleon decreases rather slowly from 
about $80\%$ at 200A MeV to still $50 \%$ at 1A GeV. As in these energies
the 'available' c.o.m. energies are well beyond typical nucleon binding
energies, dynamic mechanisms, such as cooling by adiabatic expansion, or
non-equilibrium processes must be invoked to explain such high degrees of
clusterization.\hspace{4cm}\ \ \ \ \ \ \ \ \ \ \ \ \ \ \ \ \ \ \ \ \ 

\section{Chemical temperatures}
 
\begin{figure}[!t]
\parbox{6.6cm}{
\epsfig{file=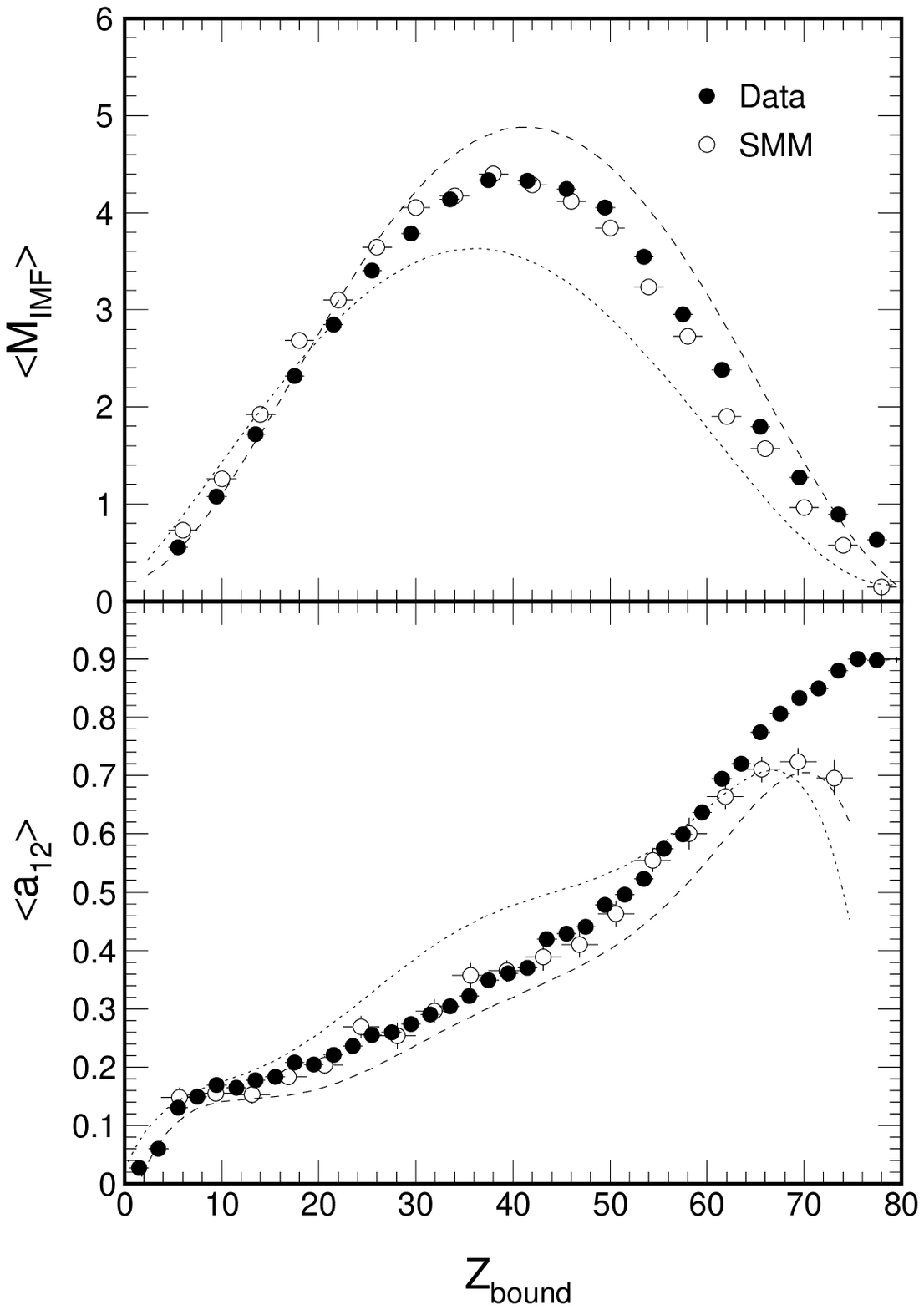,width=6.6cm}
\caption{Average intermediate mass fragment multiplicity (upper panel)
and average charge asymmetry (lower panel) as a function of $Z_{bound}$ for
experiment (full dots) and statistical multfragmentation theory (open
dots). The dotted and dashed curves show the model sensitivity to A $\pm
15\%$ variation of the excitation energy.~\protect\cite{Odeh99}} 
\label{Odeh44}
}
\hspace{6.6mm}
\parbox{6.6cm}{
\vspace{2.4cm}
\epsfig{file=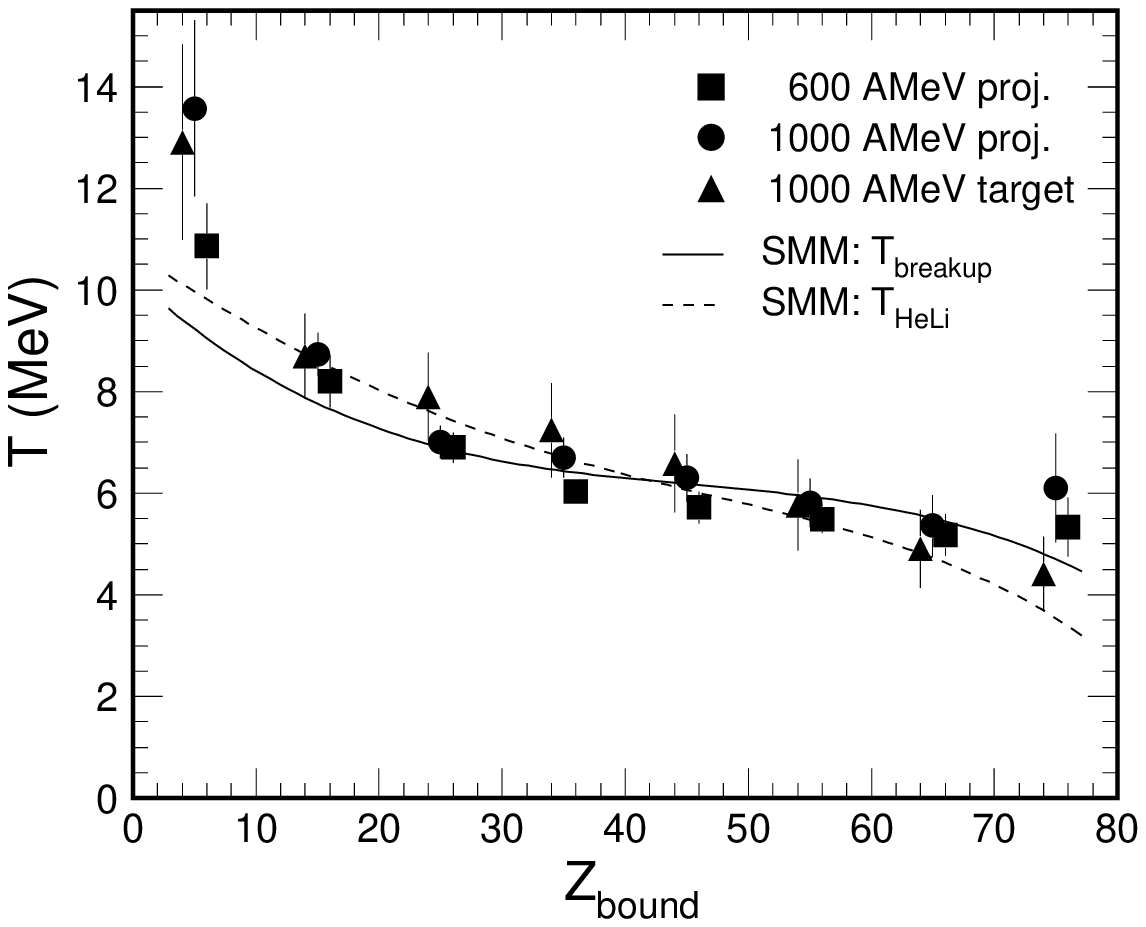,width=6.6cm}
\caption{The temperature T(HeLi) as a funtion of $Z_{bound}$.
Both projectile (600 and 1000A MeV) and target (1000A MeV) spectator data
are given. The dashed line represents T(HeLi) as determined with use of the
statistical model and the full line gives the model's breakup
 temperature.~\protect\cite{Odeh99} }  
\label{Odeh45}
}
\end{figure}

\begin{figure}[!b]
\vspace{-2cm}
\vspace{-0.5cm}
\begin{center}
\epsfig{file=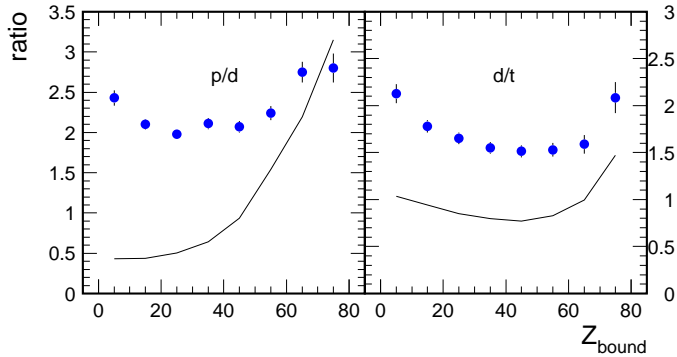,width=10cm}
\end{center}
\caption{Proton to deuteron (left) and deuteron to triton ratios as
function of $Z_{bound}$ in the reaction Au+Au at 1A GeV. The experimental
data are represented by data, statistical model calculations are given by
the smooth lines. From ref.~\protect\cite{Odeh99} }
\label{Odeh49}
\end{figure}

Frequently 'Statistical Multifragmentation Models' (SMM) are used to
compare with data under the assumption of at least chemical equilibrium.
The physics input involved are 
degrees of freedom of hot liquid drops modeled in analogy to the
Weizs\"{a}cker formula for cold nuclei,
a more or less extended list of known discrete levels and some global
level density formula to extrapolate to unresolved continuum excited
states.
Subtle sampling methods (within the grand-canonical, canonical or even
microcanonical framework) allow one then to dial typical 'events' under the
minimal assumption of basically homogeneous available phase space
population. 
Before being able to compare with experiment the 'primordial' fragment
distribution has to be corrected for later sequential decays and the
adiabatic pickup of potential field energy prevalent at the moment of
'freeze out'. (Usually only the Coulomb part is taken into account).
Comparison of one such model~\cite{Bondorf95,Botvina95} with the ALADIN
data is shown in Figs.~\ref{Odeh44}, \ref{Odeh45}, \ref{Odeh49}
 taken out of
the thesis work~\cite{Odeh99} of T.~Odeh.
Typically, one assumes freeze out at a nucleon density $\rho = 1/3
\rho_{0}$ ($\rho_{0}$ is the saturation density), and adjusts the
excitation energy (which such static theories cannot predict) until
the data are reproduced. 
As the ALADIN curve must come down to zero at the left end because
of reaching zero spectator size, as well as at the right end 
(high values of $Z_{bound}$) because
grazing collisions do not lead to any sizeable excitation, there is
sensitivity to the theory, and in particular to the assumed excitation energy,
only for intermediate $Z_{bound}$ values.
Excitation energies of about 4, resp. 10 MeV per nucleon are deduced
for spectator nuclei with masses 160, resp. 40.
Once the ALADIN curve is reproduced, 
(upper panel Fig.~\ref{Odeh44}),
one finds that  other observables
characterizing partition space are reproduced as well: see for instance
the lower panel of Fig.~\ref{Odeh44}
where the asymmetry parameter
$a_{12} = (Z_{max}-Z_{2})/(Z_{max}+Z_{2})$ is plotted ($Z_{max}$ and $Z_2$ are
the largest and next-to-largest fragment charges).

The chemical temperature deduced from the model is displayed in
Fig.~\ref{Odeh45}                
and turns out to be rather flat in the $Z_{bound}$
range where there is sensitivity to the data adjustment.
A well known method~\cite{Albergo85} to obtain density-independent
temperatures consists in evaluating double isotope yield ratios, such as
($^6$Li/$^7$Li) / ($^3$He/$^4$He). 
The deduced temperature, dubbed T(HeLi), corrected for sequential decays,
is also plotted in Fig.~\ref{Odeh45}             
together with that deduced from the
statistical model applying to it the same treatment as to the isotope data.
The remarkable consistency of these apparent temperatures has to be
confronted, however, with the failure of the model to predict correctly
the ratios of the hydrogen isotopes, Fig.~\ref{Odeh49}.
 
\begin{figure}[!h]
\parbox{6.6cm}{
\hspace{-1.1cm}
\epsfig{file=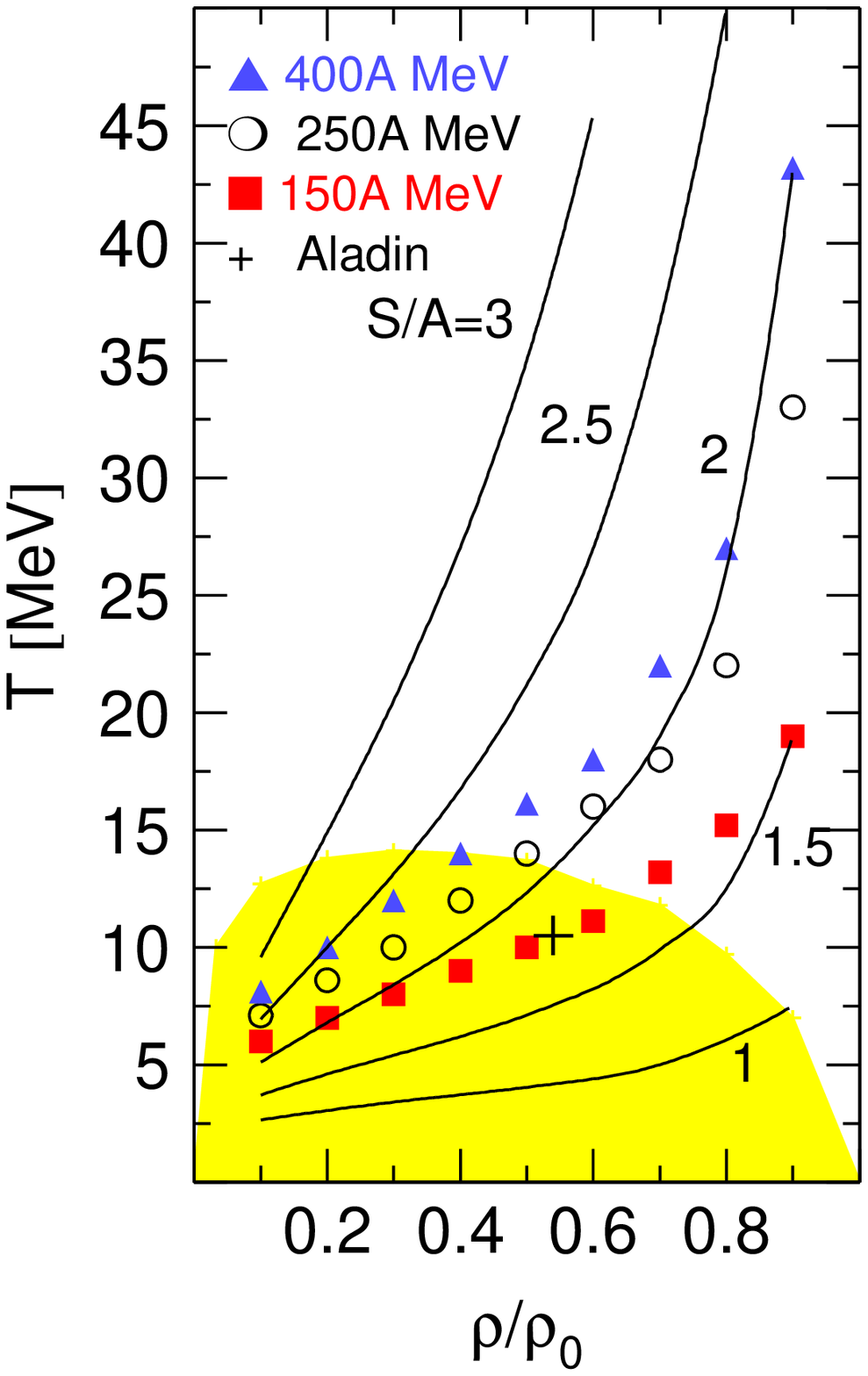,width=6.6cm}
\caption{Temperature versus freeze-out density as determined from the
comparison of the experimental charge distributions with those calculated
with QSM for Au+Au at 150A, 250A and 400A MeV. Some isentropic curves
(constant S/A) are also shown.~\protect\cite{Kuhn93} \ The shaded area
represents the spinodal region.~\protect\cite{Mueller95} \
Cross: inferred from adjacent Figure. See text.}   
\label{Kuhn}  
}
\hspace{6.6mm}
\parbox{6.6cm}{
\vspace{-1.6cm}
\hspace{-0.5cm}
\epsfig{file=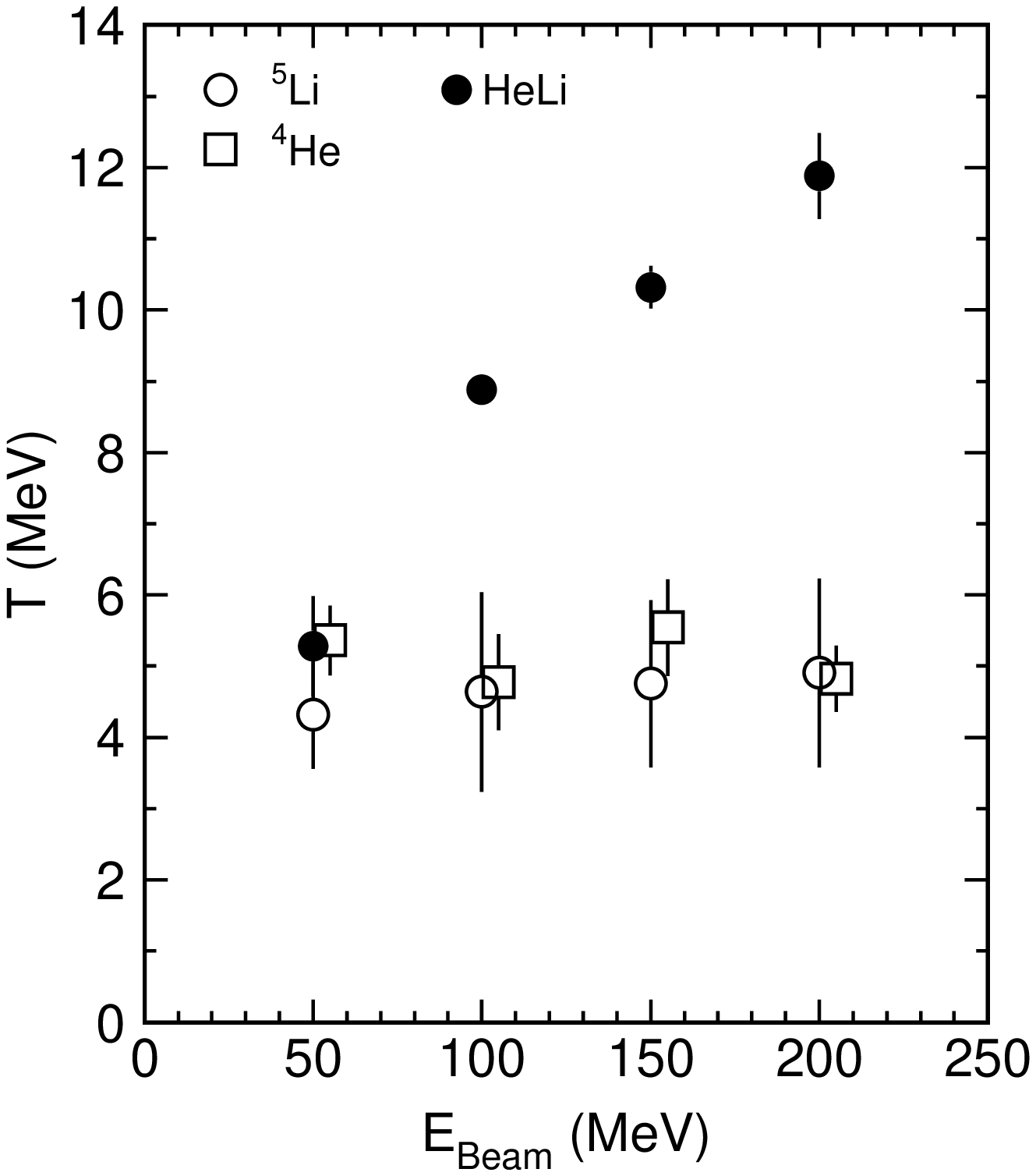,width=7.6cm}
\caption{Temperature T(HeLi) (full symbols) and excited state temperatures
 (open symbols) as a function of incident energy per nucleon in
Au on Au collisions.~\protect\cite{Serfling98} }
\label{Schere}
}
\end{figure}

\afterpage{\clearpage}
\newpage
Can the clusterization in the fireball, Figs.~\ref{zdist}, \ref{cluster},
also be understood in terms of chemical equilibrium?
In view of the simple exponential shapes of the charge distributions, it
does not come as a surprise that, given a fixed freeze-out density $\rho$,
and the freedom to adjust the overall normalization, one can find an
apparent chemical temperature $T$ that reproduces the measured slopes well.
The $(\rho , T)$ pairs compatible with the FOPI data of Fig.~\ref{zdist}
in the framework of the Quantum Statistical Model~\cite{Hahn88}, QSM, are
plotted in Fig.~\ref{Kuhn}       
adapted from ref.~\cite{Kuhn93}.
 
As the $(\rho , T)$ pairs run approximately along lines of constant
specific entropy $(S/A)$, one can use this kind of study to deduce the
latter.~\cite{Kuhn93} \
Just for the sake of orientation, the spinodal region, 
calculated~\cite{Mueller95} for
isospin symmetric matter, is also indicated in the figure.
Unless one allows for rather high freeze-out densities (such as $0.8 \rho_0
$), these apparent temperatures are rather low as we will see later. 

Fireball matter was also studied in a Au on Au experiment with the ALADIN
setup.~\cite{Serfling98} \
The temperature T(HeLi) is found to rise steadily with incident energy,
Fig.~\ref{Schere},        
and is compatible with the FOPI results in
Fig.~\ref{Kuhn}            
if the overlapping T(HeLi) point at 150A MeV
 (cross in Fig.~\ref{Kuhn})
is inserted at a density close to $0.5 \rho_0$ .

Again, a warning is given by Nature: Temperatures deduced~\cite{Serfling98}
from excited-state populations ($^5$Li, $^4$He, $^8$Be) seem to cluster
around 5 MeV and are invariant against variation of the incident energy.
See Fig.~\ref{Schere}.                  
\section{Peripheral collisions:
Kinetic energies, calorimetry}
The considerations in the previous section left out the information on
momentum space distributions.
\begin{figure}[!h]
\vspace{-1cm}
\begin{center}
\epsfig{file=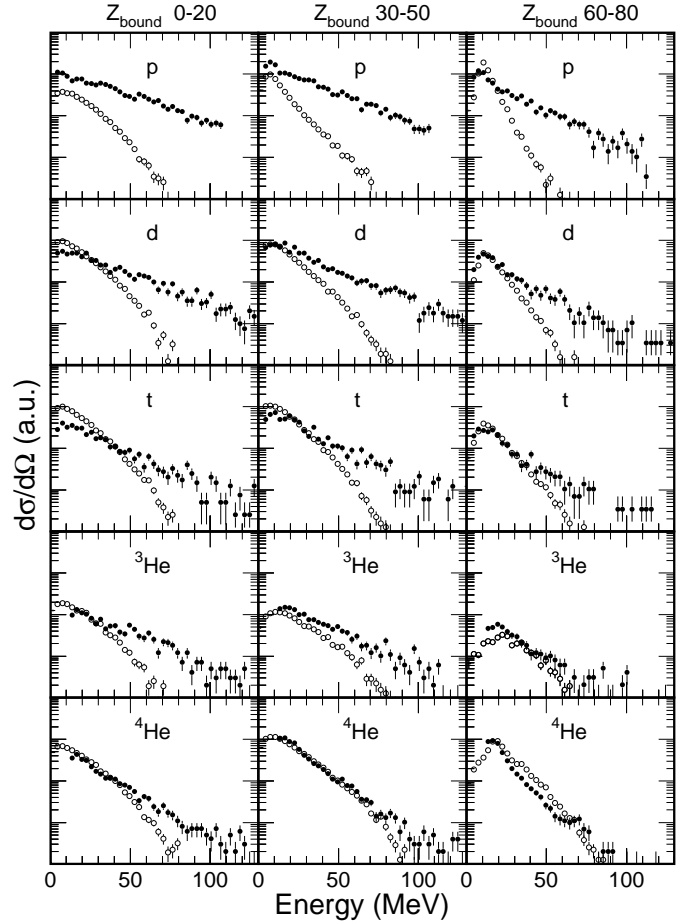,width=10cm}
\end{center}
\vspace{-0.5cm}
\caption{Energy spectra of light charged particles in three different
bins of $Z_{bound}$. Solid dots: Experimental data obtained at $150^\circ$
in the laboratory. Open symbols: SMM calculations.~\protect\cite{Odeh99}}
\label{Odeh51}
\end{figure}
 
\begin{wrapfigure}{l}{6.6cm}
\vspace{-1cm}
\epsfig{file=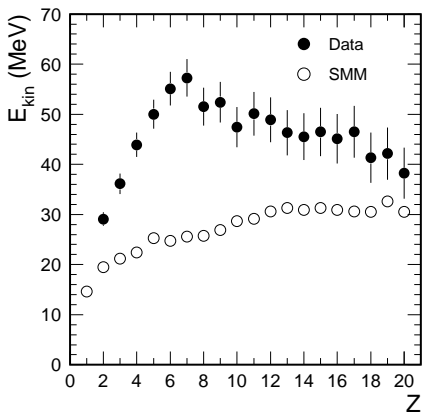,width=7cm}
\caption{Average kinetic energies of spectator fragments as function of
their charge in the reaction Au+Au at 1A GeV. Solid symbols: data,
open symbols:
SMM calculations.~\protect\cite{Odeh99}}
\label{Odeh50}
\end{wrapfigure}

\begin{figure}[!h]
\parbox{6.6cm}{
\epsfig{file=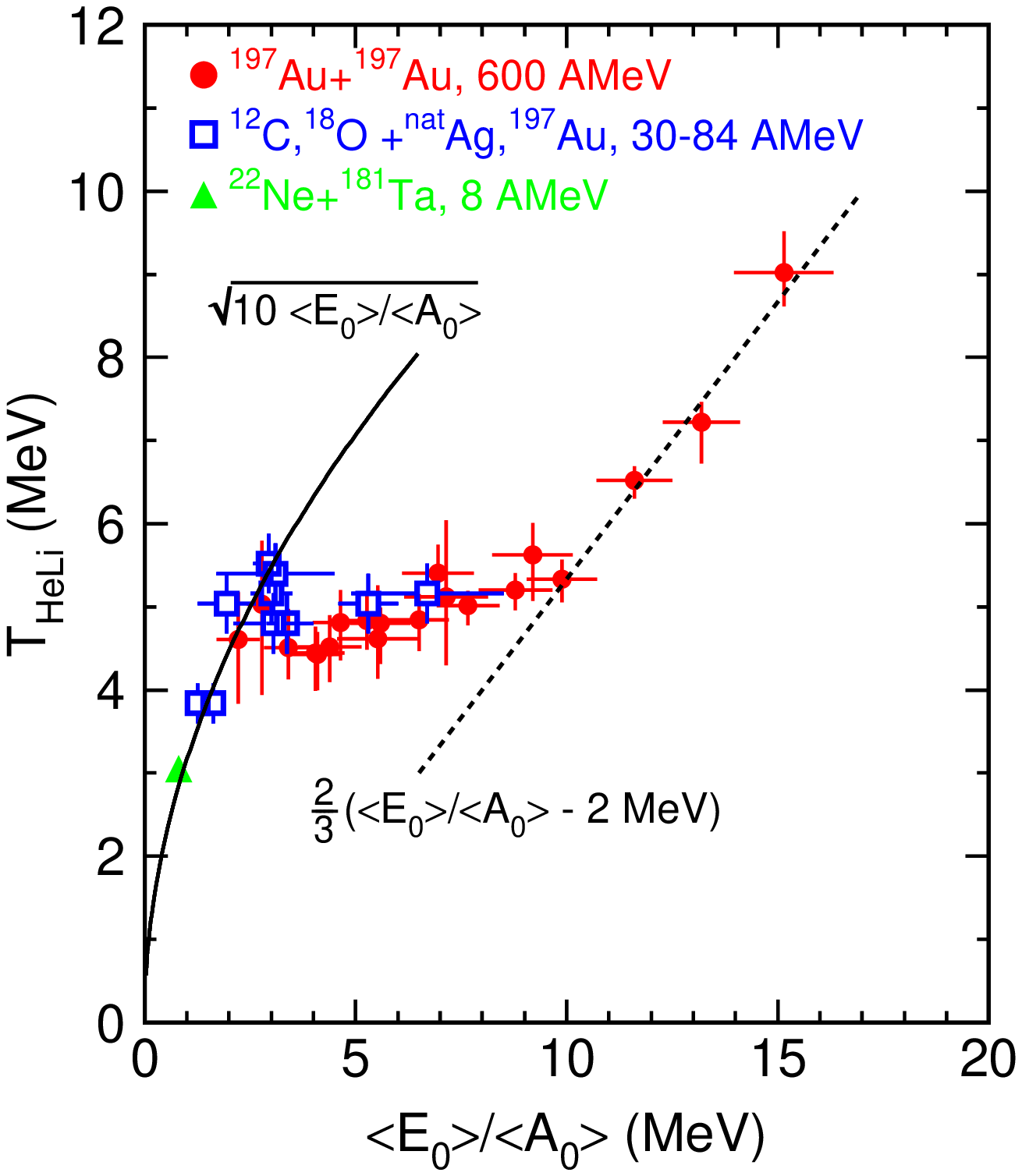,width=6.6cm}
\caption{The caloric curve showing the dependence of the isotopic
temperature on the excitation energy per nucleon.
From ref.~\protect\cite{Pochodzalla95}} 
\label{calcurve}
}
\hspace{6.6mm}
\parbox{6.6cm}{
\hspace{-0.5cm}
\epsfig{file=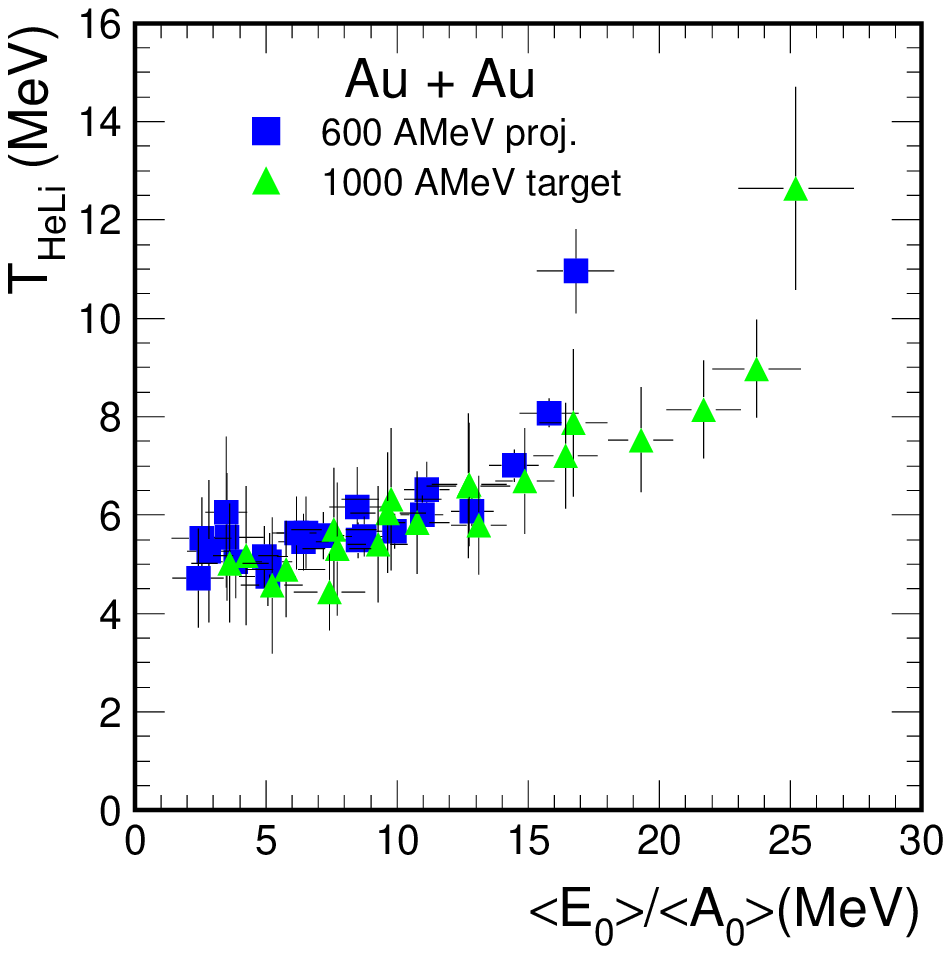,width=7.6cm}
\caption{Comparison of the caloric curves obtained for Au+Au at 600A and
1000A MeV.
From ref.~\protect\cite{Odeh99}}          
\label{calcurve1}
}
\end{figure}

\noindent
Figure~\ref{Odeh51} (light charged particles) and \ref{Odeh50}
(IMF's) summarize, for Au on Au at 1A GeV, 
the remarkable differences between the experimentally
observed kinetic energies and the predictions of the statistical approach,
that was adjusted to render correctly the yields of $Z > 2$ 
spectator fragments.
Taking a closer look at Fig.~\ref{Odeh51} (the normalization of theory
was done at the $^4$He level, lowest panels), one notices that the
deviations between theory and experiment increase from more peripheral
(large $Z_{bound}$, right column) to more central collisions (left column)
and decrease as the fragment mass is raised from $A=1$ (top row) to $A=4$
(bottom row).
The average kinetic energies of the IMF's, Fig.~\ref{Odeh50}, also
follow a remarkable pattern that differs from the statistical-model
calculation.
The latter predicts a monotonic, slowly rising trend with
nuclear charge, due to Coulomb effects.
It has been argued~\cite{Gossiaux97} that such observations give strong
evidence for a non-equilibrium situation, that can only be treated
correctly by transport model approaches, such as Quantum Molecular Dynamics
(QMD).\cite{Aichelin91} \ 
It was also proposed~\cite{Odeh99} that a sizeable part of the kinetic
energy stems from fast freeze-out of Fermionic motion.\cite{Bauer95}.
Whatever the final explanation will be, such 'non-thermal'
or 'non-classical'  phenomena make
the determination of the so-called caloric curve~\cite{Pochodzalla95} 
i.e. a plot of the temperature versus the excitation energy per nucleon,
very difficult. 
If such a curve, see Fig.~\ref{calcurve}, 
is to represent a state
property of (finite) nuclear systems, perhaps suggesting by its
plateau-like behaviour a liquid-to-gas transition, then it should be stable
against the method applied.
The major problem is to determine the abscissa which is usually obtained by
the 'calorimetric' method, consisting essentially in adding up all the
measured kinetic energies, perhaps with some high-energy tails cut off in a
more or less well justified way.
A preliminary result~\cite{Odeh99} obtained recently,
Fig.~\ref{calcurve1},  shows that the
apparent plateau width depends on the incident energy.  
 
Despite these unsolved problems in deriving a 'robust' caloric curve, it is
highly interesting that theoretical approaches using fermionic         
dynamics~\cite{Schnack97,Sugawa99} predict caloric curves that have great
similarity to the curve~\cite{Pochodzalla95} presented in
Fig.~\ref{calcurve}.

\section{Central collisions:\\
Flow and system-size dependences}

\begin{wrapfigure}{l}{6.6cm}        
\vspace{-1.5cm}
\hspace{-1.5cm}
\epsfig{file=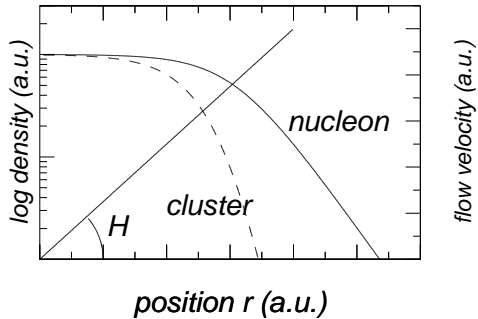,height=6.6cm}
\caption{Freeze out profile (schematic): density and flow velocity
(straight line)}
\label{profile}
\end{wrapfigure}
\begin{figure}[!t]
\begin{minipage}{7cm}
\epsfig{file=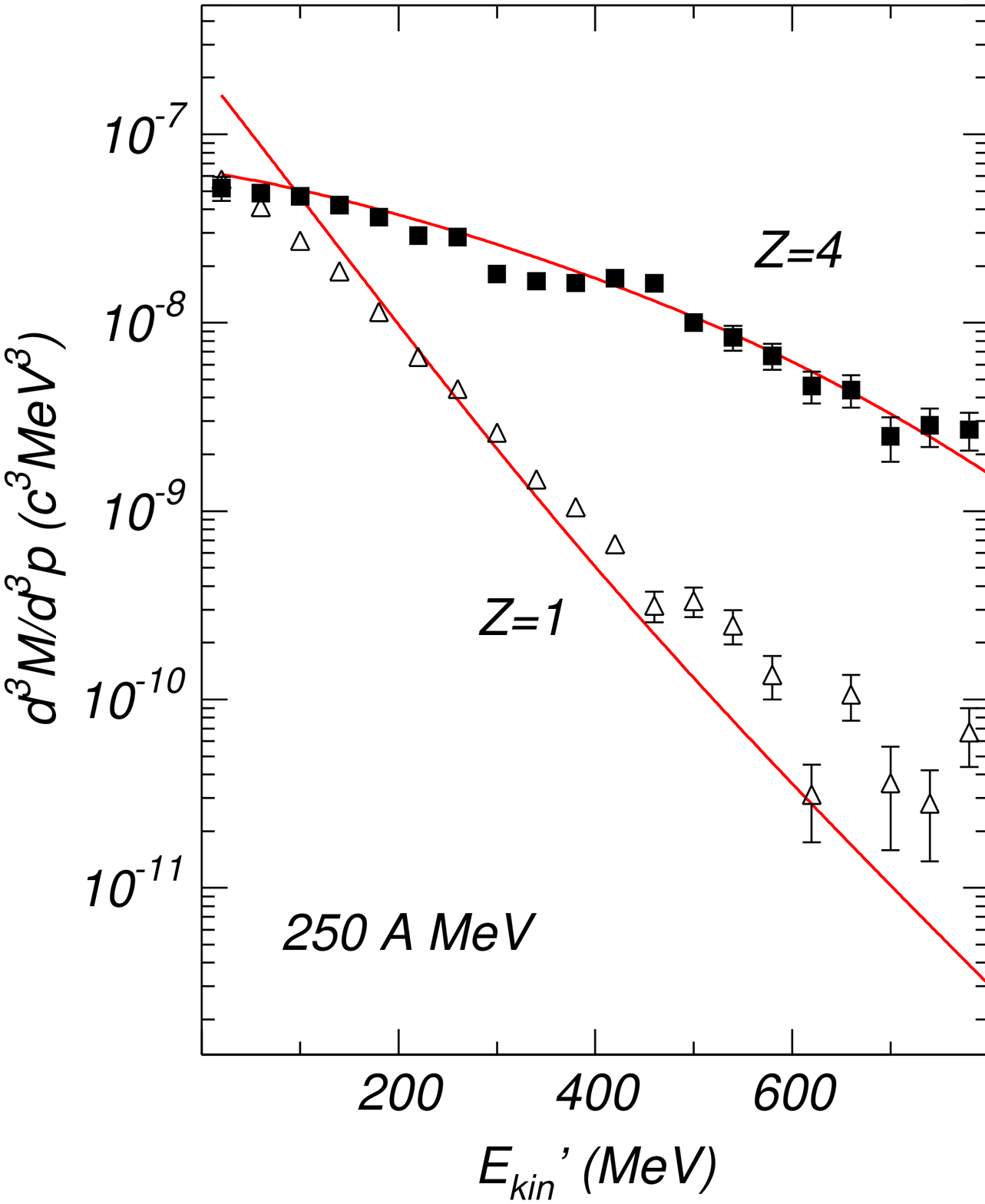,width=6.6cm}
\end{minipage}
\begin{minipage}{7cm}
\epsfig{file=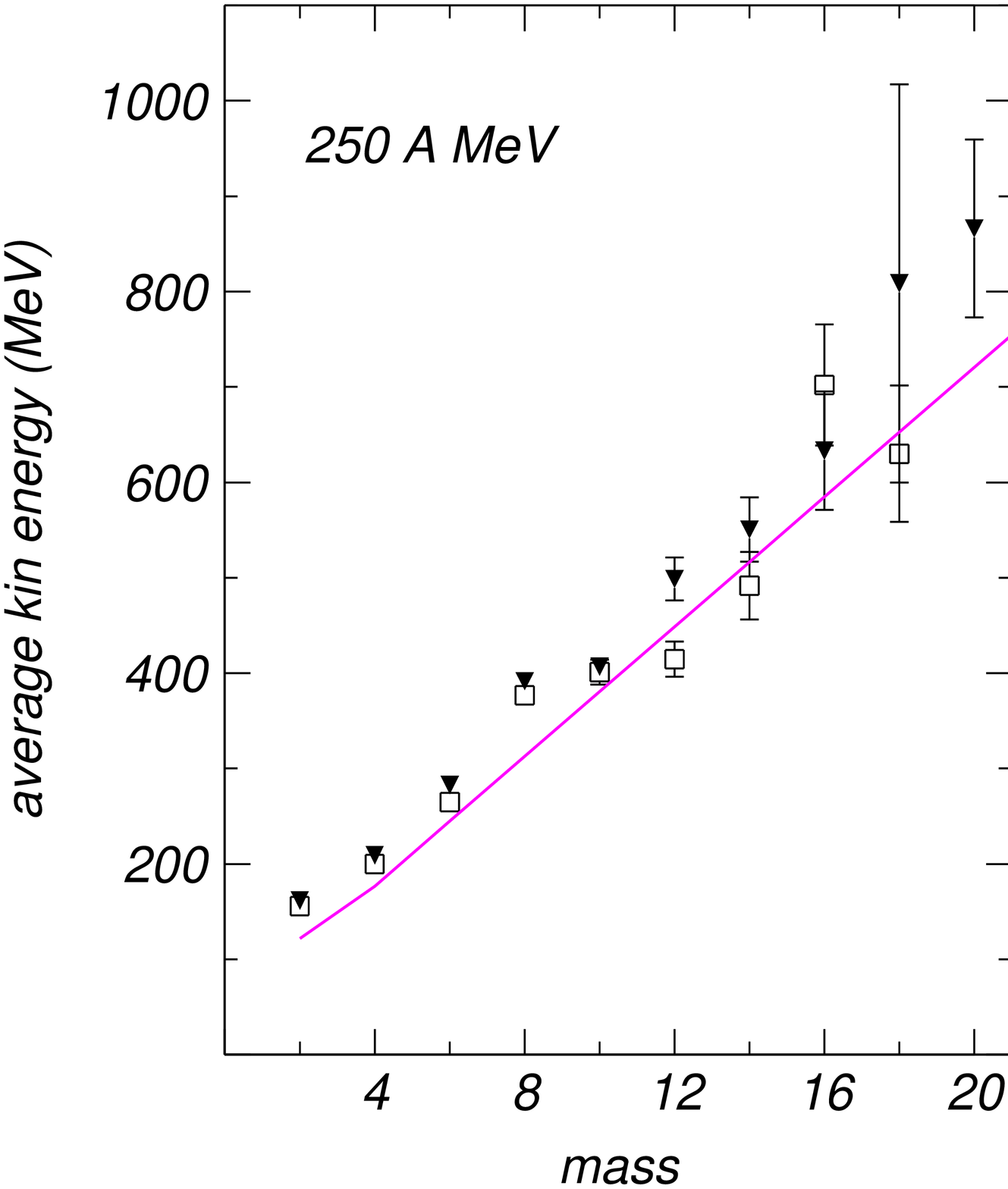,width=6.6cm}
\end{minipage}
\caption{Left panel: Kinetic energy spectra of Z=1 and Z=4 fragments in
central Au on Au collisions. Right panel: Mass dependence of average
kinetic energies. The smooth curves in both panels are blast-model fits
to the data.}
\label{blast}
\end{figure}

The fireball is not enclosed in a box (thermal model) but is expanding in
much the same way as the universe.
At freeze out the configuration of nucleons is expected to look roughly as
shown in Fig.~\ref{profile}.
There is a Woods-Saxon like density profile with a bulk
part, having subsaturation density and an extended even lower-density
surface part.
Within the bulk part the flow velocity is rising linearly.
With such a scenario, and assuming that the local temperatures are not
varying much over the configuration, one can describe kinetic energy
spectra \cite{Reisdorf97} of emitted clusterized nucleons
(Fig.~\ref{blast}, left).
A characteristic feature is that in contrast to a pure Boltzmann scenario
(no flow) one sees curved mass dependent spectra.
Those of heavier particles show a 'blueshift' in a spectacular way as they
are less sensitive to thermal fluctuations that sit on top of the flow
patterns and tend to wash out the flow features.
Another way of showing evidence for radial flow are the mass-dependent
average kinetic energies
(Fig.~\ref{blast}, right).
In a thermalized system enclosed in a box, rather than expanding, the
kinetic energies should be independent on mass.

\begin{figure}[!b]
\vspace{-2cm}
\parbox{6.6cm}{     
\hspace{-0.8cm}
\epsfig{file=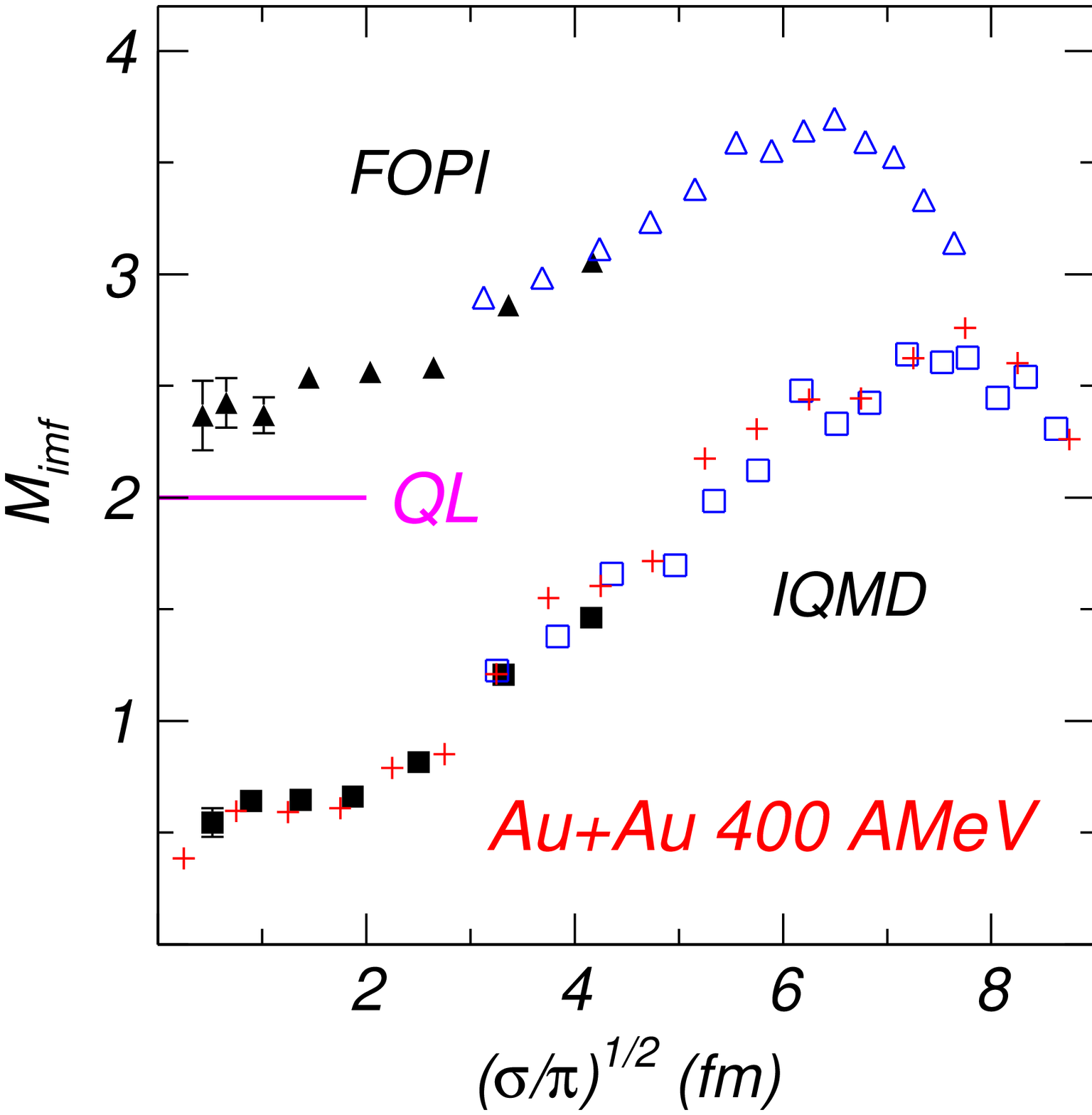,width=6.6cm}
\caption{Impact parameter dependence of IMF multiplicities in experiment
(triangles) and theory (IQMD, squares, Quantum-Langevin, QL, straight
line).}
\label{fopiqmd}
}
\hspace{6.6mm}
\parbox{6.6cm}{
\vspace{1.3cm}
\hspace{-0.8cm}
\epsfig{file=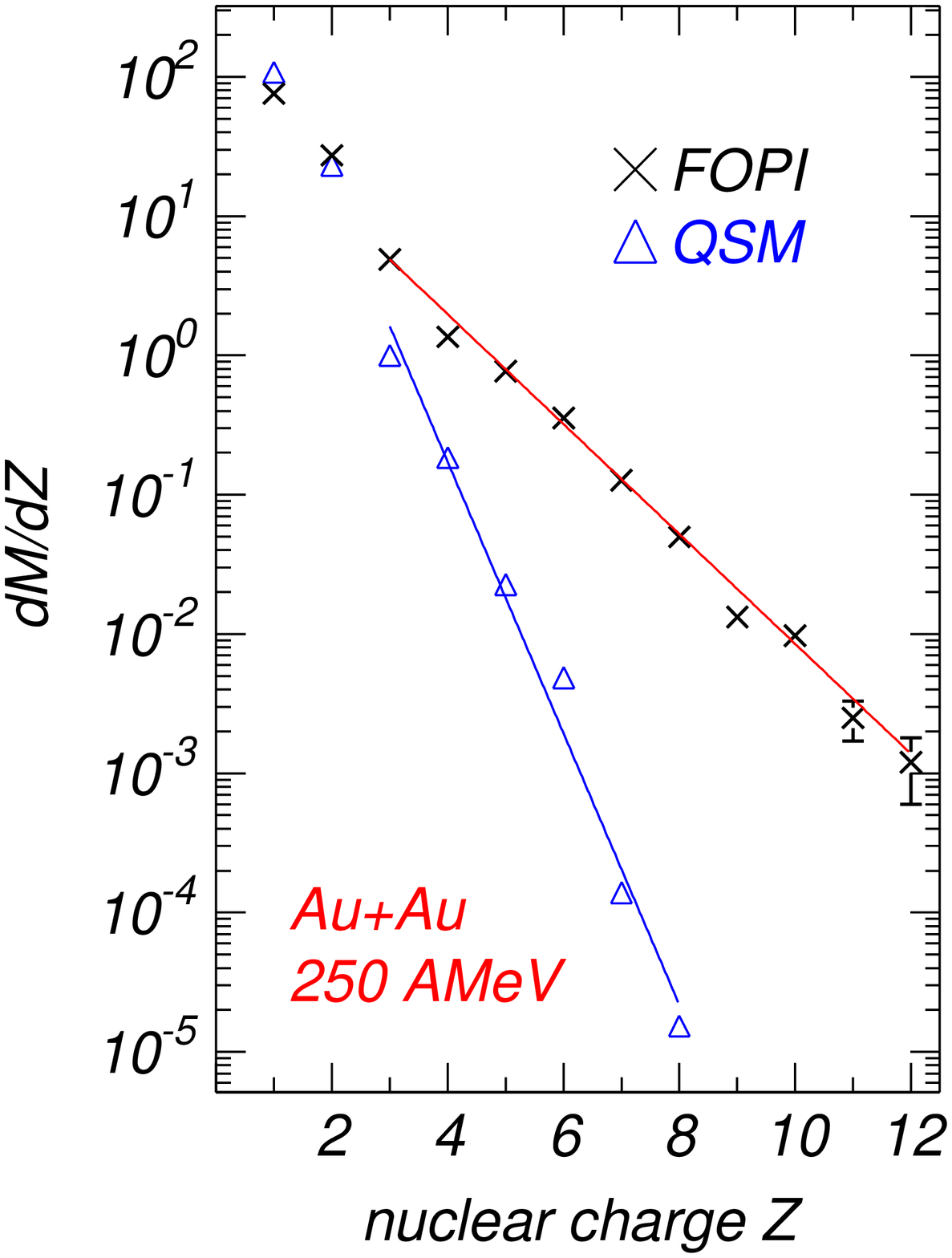,width=6.6cm}
\caption{Comparison of experimental (FOPI) charge distributions with
hydrothermal estimate}
\label{hydrothermal}
}
\end{figure}

The energy taken up by such an expansive flow was estimated to be as high
as 60$\%$.~\cite{Petrovici95,Reisdorf97}
At 250A MeV incident energy, to be specific, this still meant that about 20
MeV per nucleon (the Q-values are about $-5$A MeV) were available for
'local
use'.
It is therefore not surprising that statistical multifragmentation
calculations, allowing for the available {\em local} energy, still failed
to reproduce the measured fragment yields as illustrated 
in Fig.~\ref{hydrothermal}.
To obtain fair agreement with the data one has to assume that the primary
fragments are cold~\cite{Petrovici95},
in  contradiction to statistical expectations.

Transition to transport models, not limited by the local equilibrium
assumption, seems mandatory. 
In the past the IQMD model calculations \cite{Aichelin91,Hartnack93} have been
unable, though, to reproduce the observed IMF multiplicity, 
see Fig.~\ref{fopiqmd}.
Modern developments are promising:
a new cluster finding algorithm~\cite{Nebauer99}
seems to be successful.



One has tried to overcome the lack of quantum fluctuations in these
quasi-classical simulations by introducing either so-called
Quantum-Langevin methods \cite{Ohnishi97} or by using Antisymmetrized
Molecular Dynamics \cite{Feldmeier90,Ono99}.

In the following we discuss new, partially preliminary results obtained by
the FOPI Collaboration, that shed additional light onto this complex
subject.

\begin{figure}[!b]
\vspace{-1cm}
\parbox{6.6cm}{     
\hspace{-0.7cm}
\epsfig{file=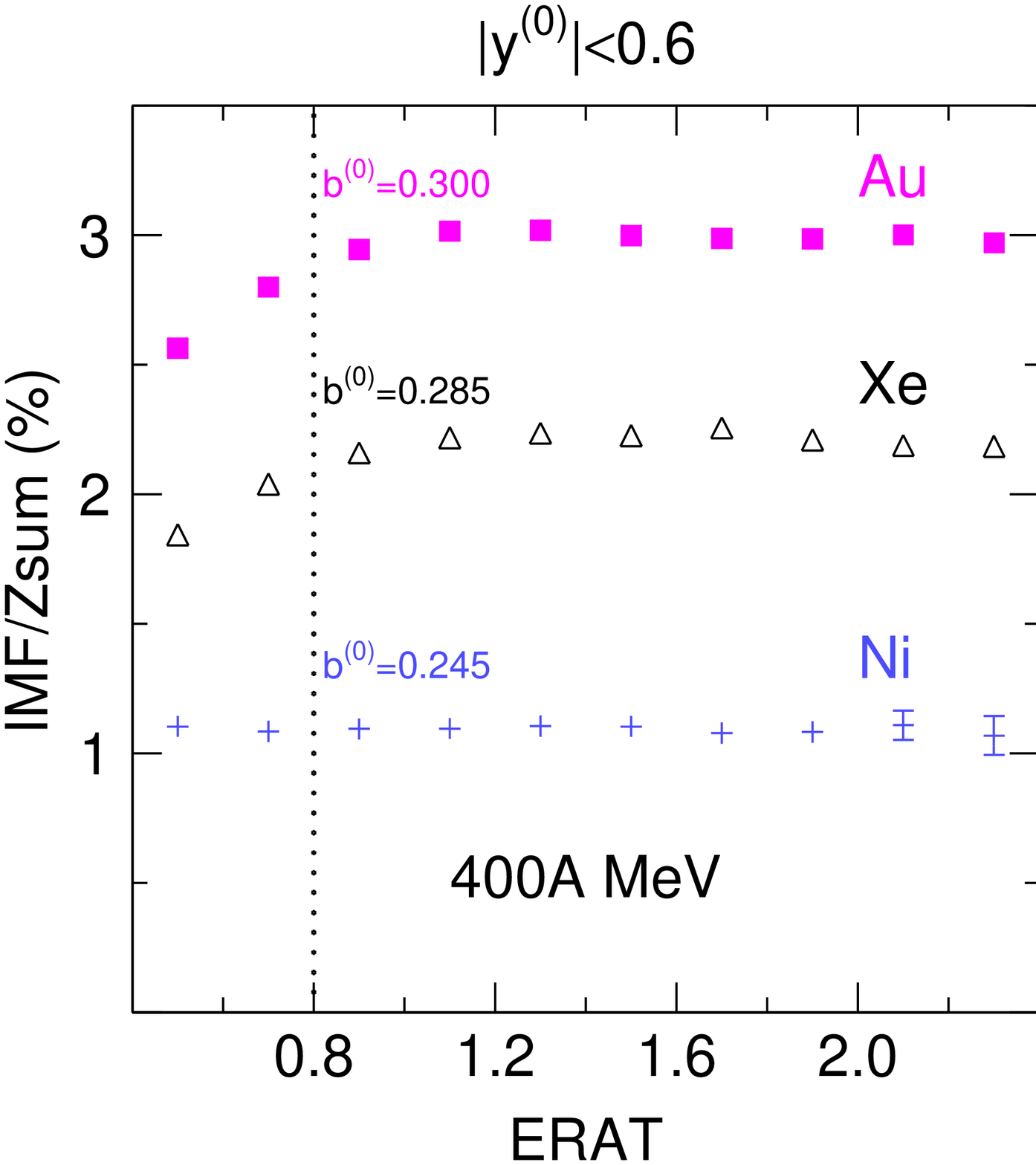,width=6.6cm}
\caption{Normalized IMF ($Z > 2$) production in various indicated 
symmetric systems. The abscissa ERAT is the ratio of total transverse to
longitudinal kinetic energy.
Values in excess of 0.8 correspond to collisions with reduced impact
parameter below 0.3.
} 
\label{mimf400}
}
\hspace{6.6mm}
\parbox{6.6cm}{
\vspace{-0.5cm}
\hspace{-0.7cm}
\epsfig{file=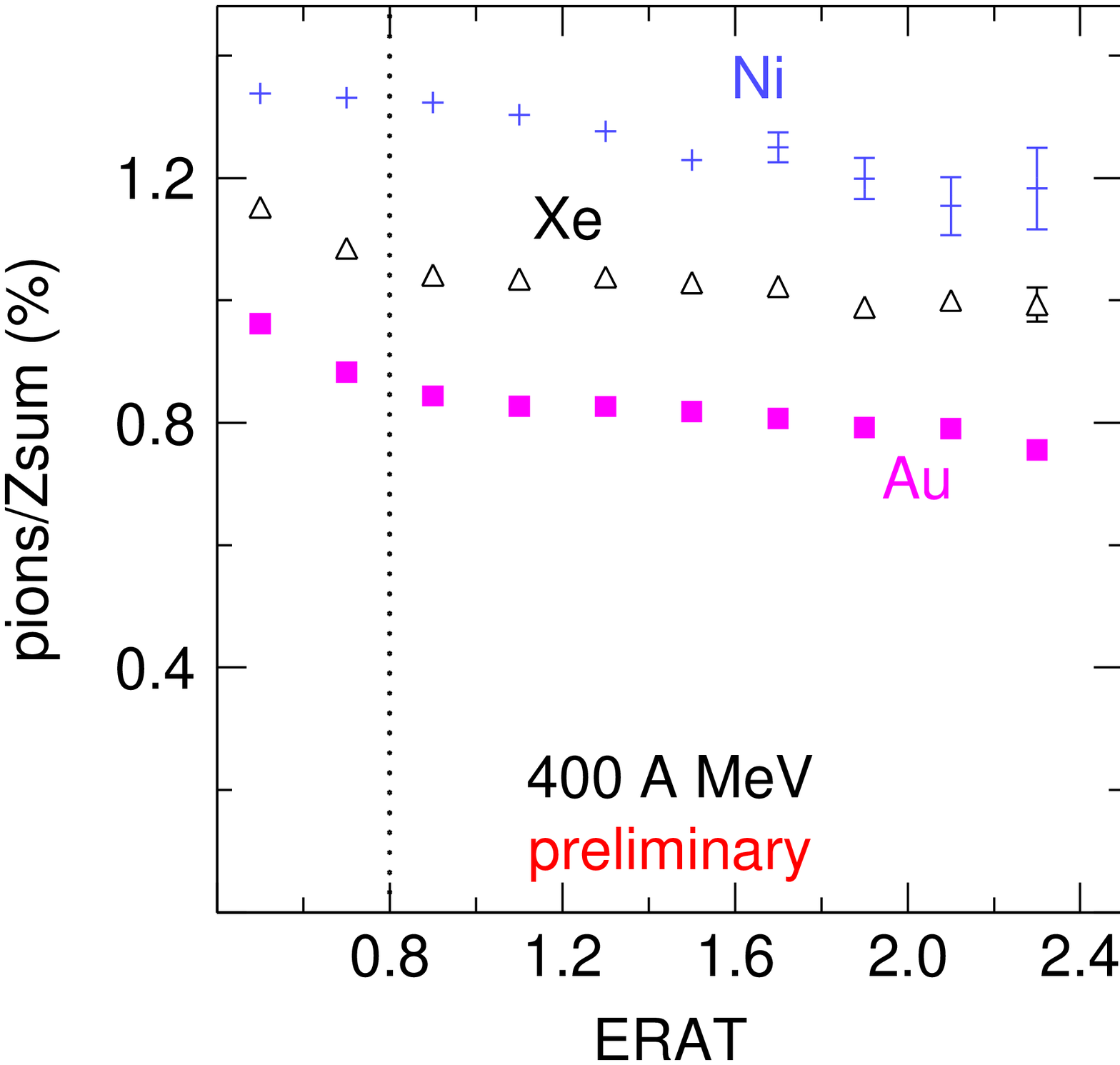,width=6.6cm}
\caption{Same as adjacent figure, but for pions}
\label{pion400}
}
\end{figure}

It is of high interest to study system size dependences.
Are the fireball data as 'universal' as the 'spectator' data are?
We have studied the number of emitted intermediate-mass fragments and the pion
multiplicities varying the system size from $^{58}$Ni+$^{58}$Ni, via 
$^{129}$Xe+CsI to Au+Au, see Figs.~\ref{mimf400} and \ref{pion400}.
Roughly, normalizing to a total of hundred observed baryonic charges,
we find that three IMF are emitted in Au on Au reactions, two IMF in Xe+CsI
and one IMF in Ni on Ni reactions.
The order is inversed for pion emission: relatively more pions are emitted
from the lighter Ni+Ni system (Fig.~\ref{pion400}).
Using thermal language, one might say that heavier systems appear to be
cooler at freeze-out, perhaps favoured by a more substantial bulk
expansion.

\begin{figure}[!h]
\vspace{-2cm}
\begin{center}
\epsfig{file=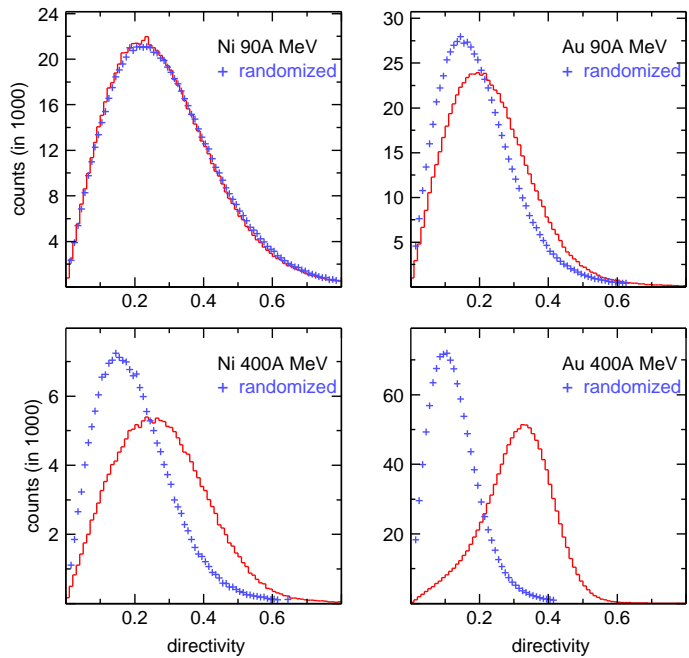,width=11cm}
\end{center}
\caption{Directivity distributions in various indicated systems at the same
normalized centrality. The signal (histogram) is compared to distributions
from azimuthally randomized events.}
\label{dir}    
\end{figure}

This seems to be corroborated by flow studies.
Coming back to the directivity $D$ introduced in section~2 to sort out
very central collisions, we compare in Fig.~\ref{dir} directivity distributions
measured for Ni+Ni and Au+Au at 90 and 400A MeV.
The reduced impact parameter range is $0 < b^{(0)} < 0.4$ (touching
configurations corresponding to $b^{(0)}=1$).
The (two-dimensional) random walk distributions of $D$ (crosses),
obtained by randomizing the azimuthal
emission angles in each event, are compared with the actual distributions
(histograms).

One finds that the random distributions of this scale-invariant observable
are broader for lighter than heavier systems (i.e. the {\em relative}
fluctuations are larger), see the top two panels in Fig.~\ref{dir},
and narrow down
at higher energies because of increasing particle multiplicities
(compare the top with the bottom panels).
Comparing with the actual distributions one sees that

\begin{wrapfigure}{l}{6.6cm}
\hspace{-0.5cm}
\epsfig{file=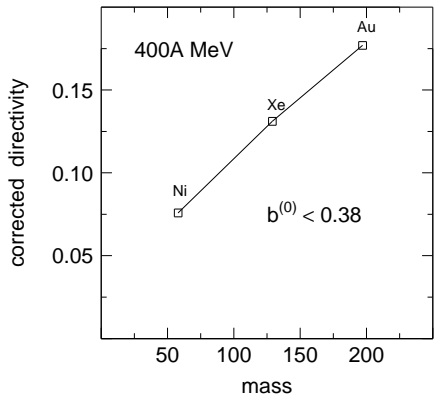,width=6.6cm}
\caption{Corrected directictivity versus projectile mass in symmetric
systems at 400A MeV beam energy} 
\label{dir2}   
\end{wrapfigure}

1) At 90A MeV there is no net effect for the Ni system, while a
statistically significant signal is seen for Au. \\
2) At 400A MeV sideflow has increased substantially in both systems.\\
3) The sideflow,  Fig.~\ref{dir2}, after correction for the
finite number fluctuation, shows a dramatic system-size dependence.

We are currently also studying other aspects of flow, such as radial flow
and out-of-plane flow as a function of system size.
It is intriguing to connect information on flow with the probability to
form clusters.
The failure of the hydrothermal attempt to understand the degree of
clusterization, Fig.~\ref{hydrothermal},           
suggests that rapid expansion physics may
well require the consideration of fast non-equilibrium mechanisms, as
proposed elsewhere~\cite{Holian88} in a different context.
The very low 'temperatures' suggested by the
analyses~\cite{Kuhn93,Serfling98}, see Figs.~\ref{Kuhn},~\ref{Schere},                        
would then be an artifact of an inadequate assumption (equilibrium).

\section{Non-equilibrium:\\
isospin tracer method}
Valuable insight into the question of equilibration can be gained by
studying systems, that, while mass symmetric, are isospin asymmetric.
We have performed an 'isopspin tracer experiment' using four mass 96+96
systems Ru+Ru, Zr+Zr, Ru+Zr and Zr+Ru at 400A MeV.
Naively, if Ru was to impinge on Zr, then, after the collision and
in the case of complete transparency,
we should be able to count 44
protons (Ru) in the forward rapidity hemisphere, and we should find the 40
protons of Zr at backward rapidities.
In the extreme limit of a {\em collective}
'rebound' (predicted by one-fluid
non-viscous hydrodynamics, as we shall see), just the opposite would be
observed, while 'complete mixing' would show up as a 'white' distribution.

\begin{figure}[!t]
\parbox{6.6cm}{
\hspace{-0.5cm}
\epsfig{file=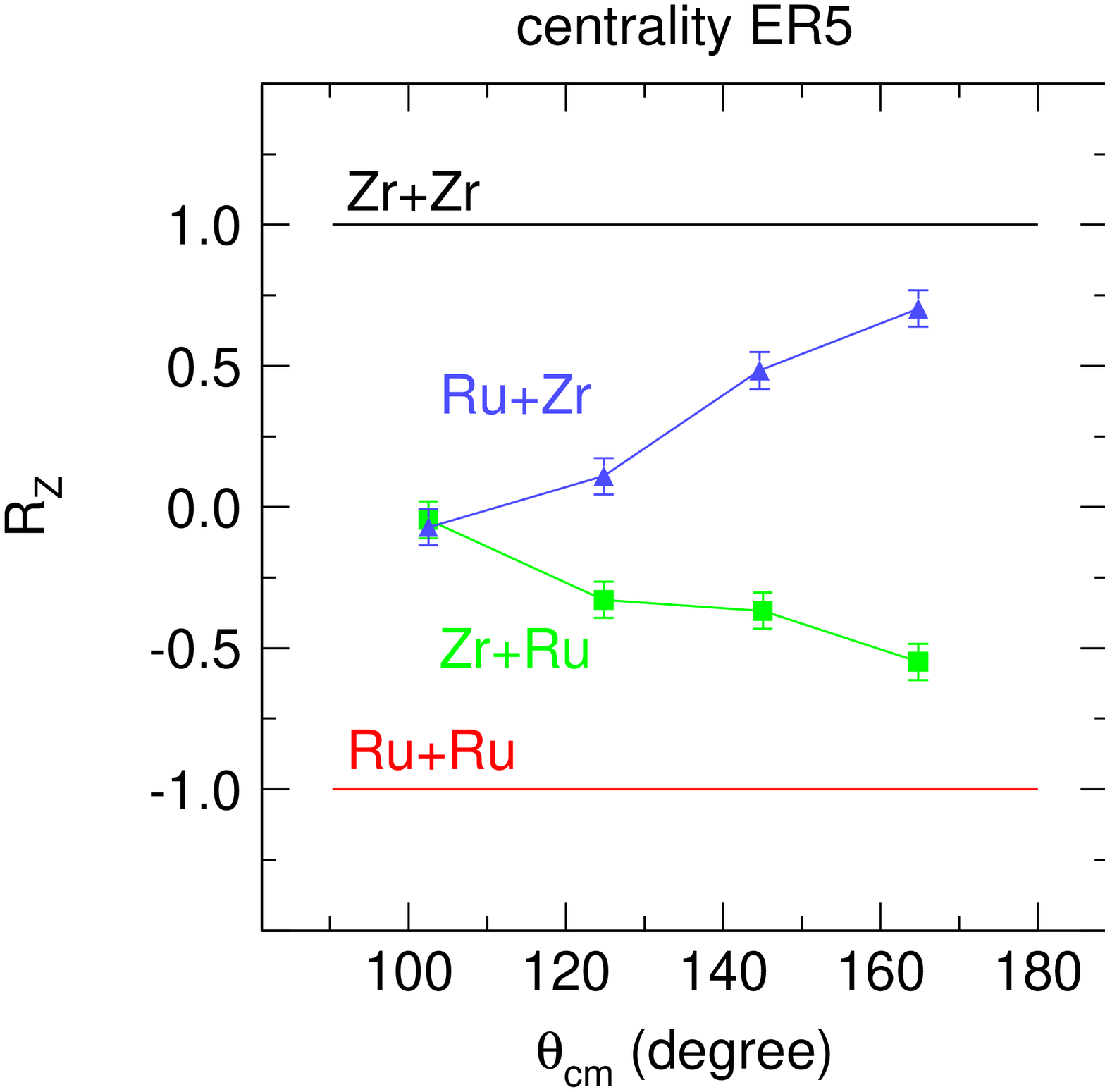,width=6.6cm}
\caption{Proton counting as a function of polar angle in four mass-96
on mass-96 systems at 400A MeV (see text).}
\label{rz}
}
\hspace{6.6mm}
\parbox{6.6cm}{         
\hspace{-0.5cm}
\epsfig{file=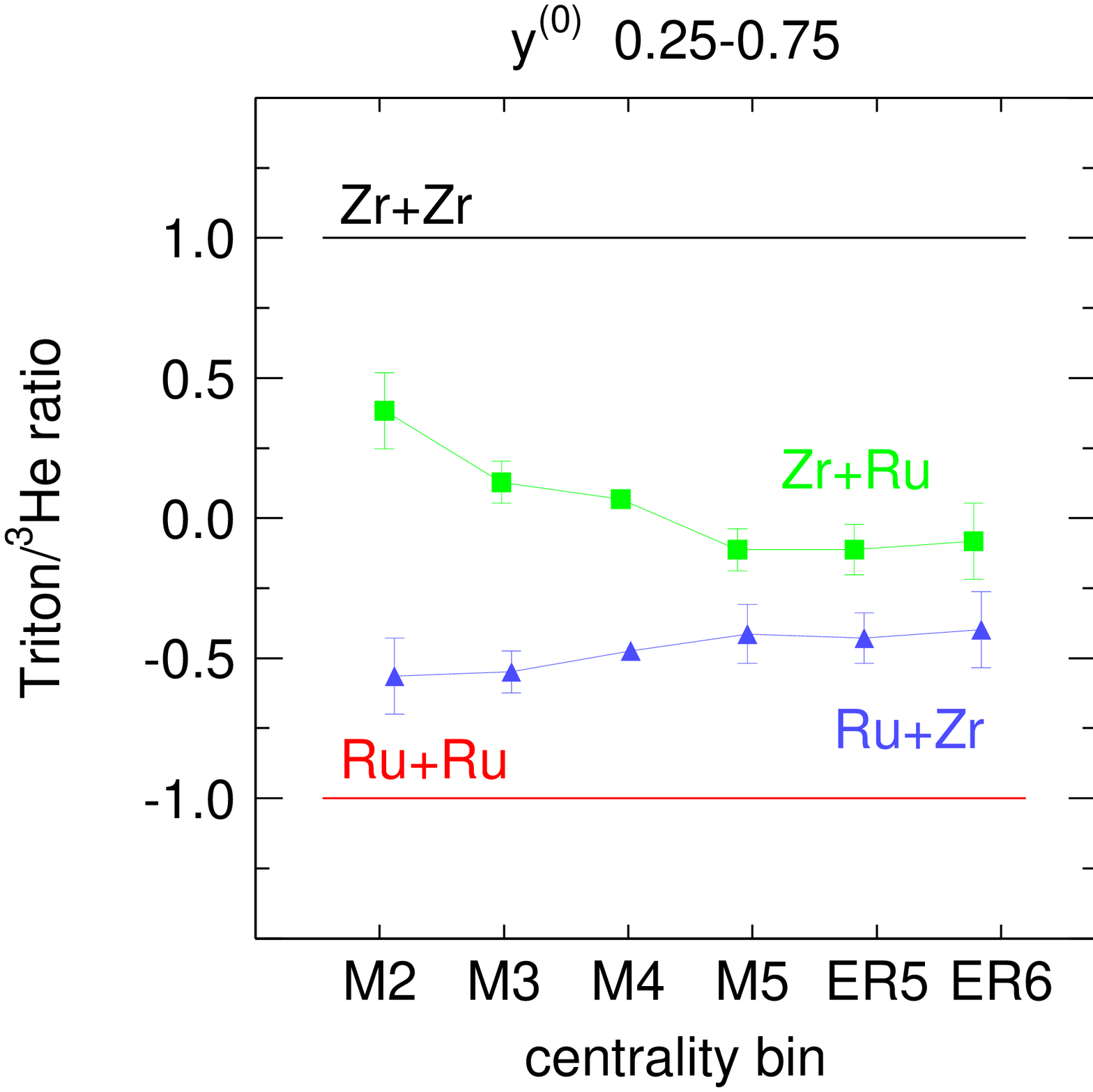,width=6.6cm}
\caption{Normalized triton/$^3$He ratios as a function of centrality}  
\label{fg:t3he}
}
\end{figure}

A more complete description of this experiment is given in
reference.~\cite{Rami00} \ 
In the studied reactions, if central collisions are selected,
the transverse-momentum vs rapidity distributions of emitted protons
indicate the apparent dominance of {\em one} nearly isotropic source
located at midrapidity.
Fig.~33 shows for very central collisions the result of 'proton-counting'
(including those in deuterons) for the four systems as a function of the 
center-of-mass polar angle.
To quantify conveniently the 'degree of mixing' we first study the
symmetric systems Ru+Ru and Zr+Zr and renormalize in each angular bin the
observable to $+1$ (Zr), resp. $-1$ (Ru).
Using this normalization, we then study the asymmetric systems Ru
$\rightarrow$ Zr and the 'inverse' reaction Zr $\rightarrow$ Ru.
As can be seen, the two asymmetric systems converge to the expected mixed
value (zero) only near $90^{\circ}$, where they {\em must} converge. 
As soon as one leaves the $90^{\circ}$ interval, the proton counting
observable tends to the value expected for transparency, the sign being
opposite for the Ru+Zr and Zr+Ru systems, as it should be for internal
consistency.


An alternative way of demonstrating this partial transparency is to use
instead the triton to $^{3}$He ratio (Fig.~34).
Again one proceeds by looking at the symmetric system first, normalizes the
difference to $\pm 1$ and then studies the asymmetric systems.
This time we fix the phase-space region: we look at to $y^{(0)}=0.25-0.75$.
We now vary the centrality from impact parameters below 1 fm (ER6) to about
8 fm (M2).
Note again that effects can only be expected away from midrapidity.
We find that our second observable again indicates partial transparency,
although the effect is least pronounced for the most central collisions.
One can show by the way \cite{Schauen} that the 'fully mixed' value should
be somewhat below the zero line in Fig.~34,   near (-0.20).
 
If one uses the proton-counting method as a function of rapidity, rather
than polar angle as was done in Fig.~33, one can actually decompose
the rapidity distribution in a projectile and a target 'remnant', as
shown in Fig.~35.

\begin{figure}[!t]
\parbox{6.6cm}{
\hspace{-0.7cm}
\epsfig{file=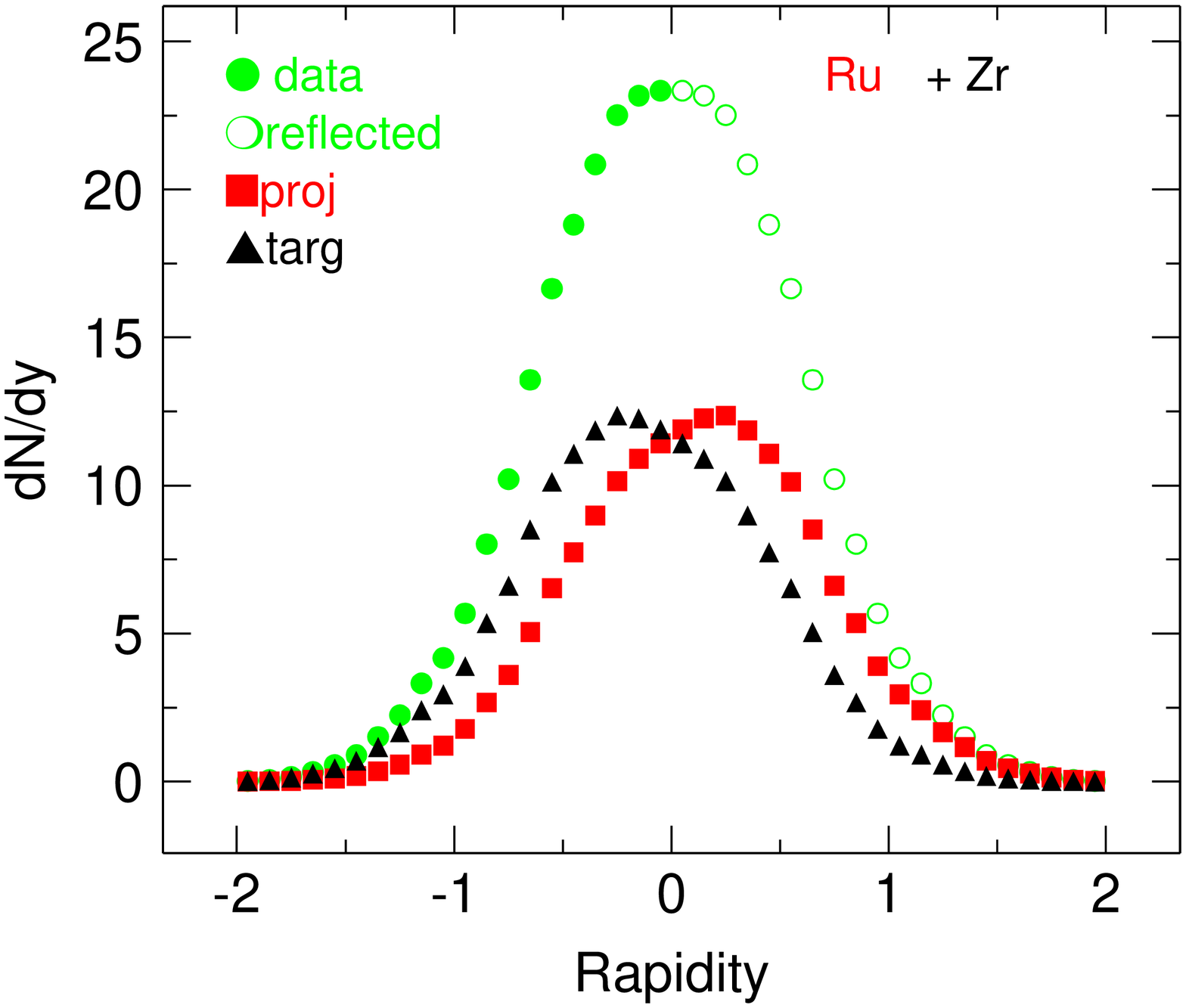,width=6.6cm}
\caption{Decomposition of the rapidity distribution in central collisions
of $^{96}$Ru + $^{96}$Zr at 400A MeV. After ref.~\protect\cite{Rami00}}
\label{nzy}      
}
\hspace{6.6mm}
\parbox{6.6cm}{       
\hspace{-0.7cm}
\epsfig{file=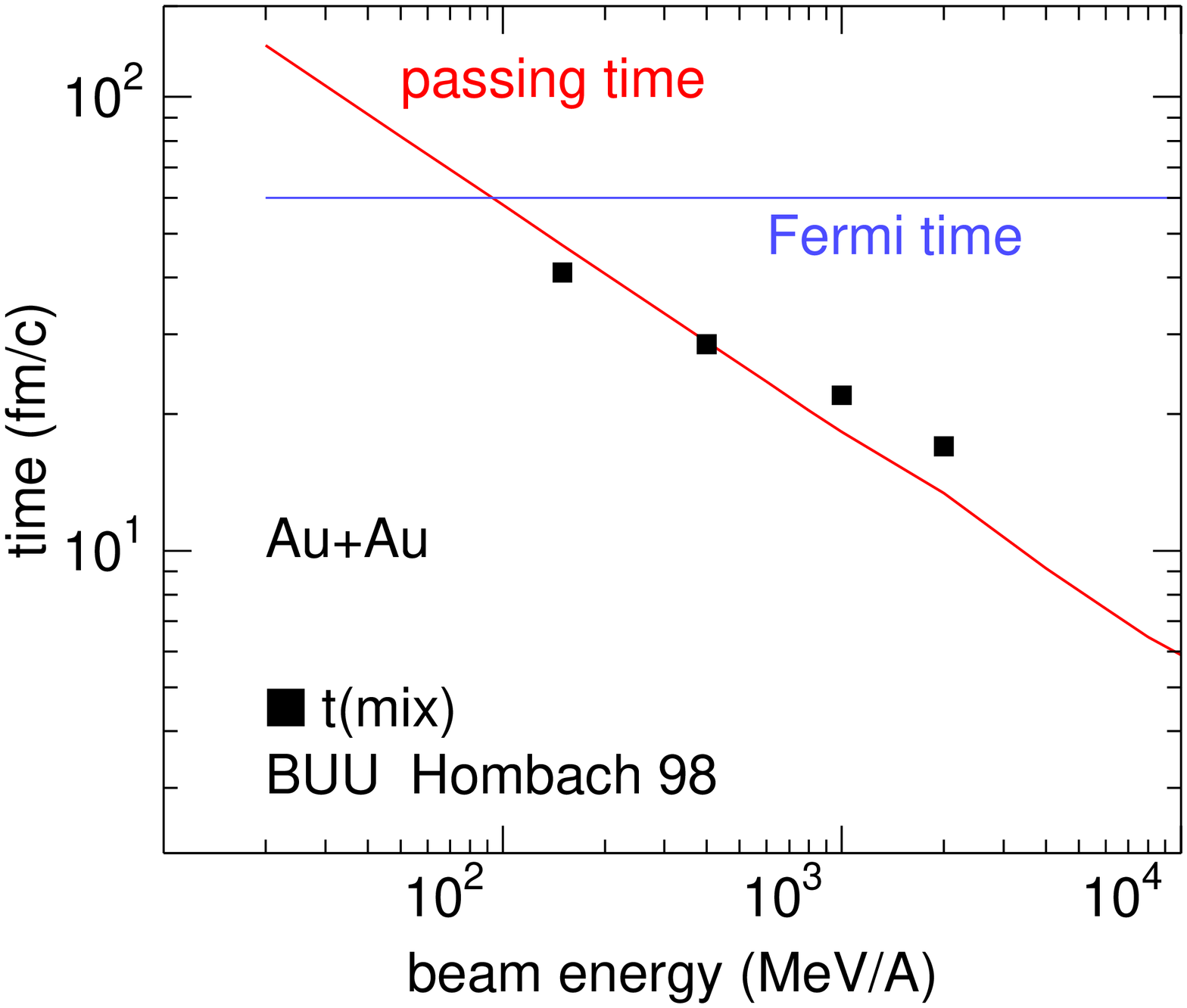,width=6.6cm}
\caption{Passing time, Fermi time and mixing time (symbols) as a function
of beam energy}
\label{timescales}
}
\end{figure}

What can we conclude?
We believe these types of data are important for at least two reasons.\\
1) it is good to have a signature of partial memory of the past: this puts a
constraint on dynamic models; these models must show that they can
realistically cope with the observed partial transparency, independently
whether its origin is connected with the surface corona and possibly
even with insufficient stopping volume. 
Reactions with heavy ions {\em always} concern finite sizes and corona
effects.\\
2) If the system is not fully equilibrated at the {\em end} of the evolution,
we expect it to be even less equilibrated at earlier stages, i.e. at
maximum compression.
Hence the pressure expected from the nuclear equation of state (EOS) is not
fully developed, which in turn will lead to decreased flow at the end of
the evolution.
 {\em Transparency may mock up a 'soft' EOS.}
Transport models used to extract information on the EOS must therefore
correctly reproduce the observed partial transparency (be it bulk and/or
surface transparency).

Clearly, corona effects must be one of the reasons for observing partial
transparency, although they are subtle: for the most central collisions
they do not appear as an obvious structure close to target or projectile
rapidity (see Fig.~35).
Various transport calculations shedding light on the mixing problem have
already been done.
In Fig.~36 we show results deduced from recent work of Hombach et
al.\cite{Hombach99}.
Using a periodic box condition the authors have determined the
equilibration time for nucleon mixing in the SIS energy range.
They find times on the order of 40 to 10 fm/c decreasing with increasing
incident energy (see the Figure).

\begin{figure}[!t]
\begin{center}
\epsfig{file=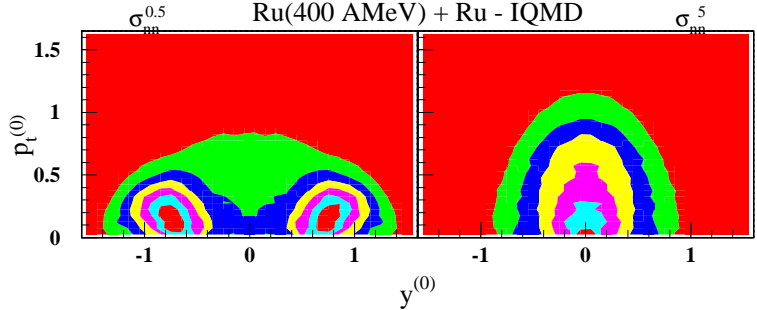,width=11cm}
\end{center}
\caption{Momentum space distributions in a transparency and a hydro-shock
scenario.}
\label{iqmd}
\end{figure}


Comparing this with the passing time $(2R/u)$ ($R$ projectile radius, $u$
four-velocity), and the Fermi-time (typical time for a nucleon to cross the
system due to its Fermi-motion), we can conclude from this study that the
passing time might be somewhat too short for equilibration below 500A
MeV. 
The relative transparency at
lower energies is intimately connected with the importance of Pauli
blocking and the still frozen nucleonic degrees of freedom.
(The mobility implied by the Fermi time is important for the communication
between 'participants' and 'spectators') .

\begin{wrapfigure}{l}{6.6cm}
\hspace{0.5cm}
\epsfig{file=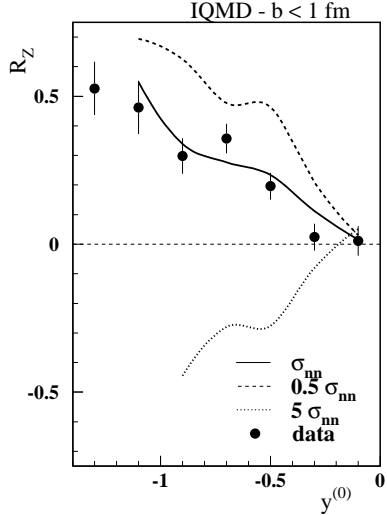,width=5.0cm}
\caption{Comparison of the proton counting observable with various
IQMD simulations}
\label{rzqmd}
\end{wrapfigure}
Transparency is sensitive to the medium-modified nucleon-nucleon elastic
scattering cross sections $\sigma_{nn}$.
Fig.~37 shows momentum-space distributions for Ru+Ru at 400 AMeV obtained
using the IQMD transport code \cite{Hartnack93} with two extreme, but
instructive, assumptions: for the left panel we used \cite{Schauen}
$0.5*\sigma_{nn}$ and for the right panel $5*\sigma_{nn}$ ($\sigma_{nn}$ are
free-scattering cross sections from the literature).
We note that the code of course applies the usual Pauli-blocking weights in
the collision term \cite{Aichelin91}.
This takes care of some, but not all in-medium effects.
The two (impact parameter zero) cases illustrate the transparent
and the 'hydro-shock' alternatives.
In the former case one still sees 'spectator' sources, in contrast to
experiment.
In the latter case, almost instant equilibration leads to highly oblate
event shapes with increased particle emission at $90^{\circ}$ (c.o.m.):
the true 'squeeze-out' predicted in the early seventies \cite{Scheid74}.
The consequences for our proton counting observable are seen in Fig.~38.
In particular the hydro-shock scenario leads to a rebound, the opposite of
what we observe.
The data are in between the two scenarios (although closer to the
'transparent' scenario) and are in impressive agreement with IQMD
predictions using the nominal $\sigma_{nn}$ (solid curve).
It is important to realize that only a {\em fraction} of the flow-creating
pressure expected from ideal hydrodynamics is achieved in a heavy ion
collision.

\section{Conclusion}
We have presented multifragmentation data from heavy ion reactions at
incident energies varying from 0.1A to 1A GeV.
Two powerful and to some degree complementary experimental setups, ALADIN
and FOPI, allowed us to study the complex decay of both 'spectator' and
'fireball' or 'participant' matter.
It appears that rather different mechanisms govern these multiparticle
events.
While a remarkable 'universality', i.e. invariance under variation of the
system size and of the incident energy, was seen to characterize the decay
of the intermediate mass fragment source in the case of spectator matter,
we saw that in contrast, the decay of rapidly expanding fireball matter
showed pronounced system-size dependence, that could be traced, at least in
part, to the occurrence of significant size-dependent flow.

Purely statistical concepts, comparatively successful in explaining the
yields of particles with $Z > 2$, fail when it comes to explaining light
particle yields, and above all when momentum space distributions are to be
understood. Even the 'universal' spectator decays are accompanied by an
apparently fast removal of some of the available energy primarily by single
nucleons and show a non-trivial evolution of average kinetic energies with
the nuclear charge, implying the presence of either flow (sideways bounce
or radial) or fast freeze-out of Fermionic motion, possibly due to
large scale fluctuations caused by the system passing through the spinodal
region, or some entrance-channel memory effects.

Such phenomena must be understood before we can think of deriving a
'robust' caloric curve for finite nuclei, that we eventually could
extrapolate to infinite nuclear matter.

It seems also urgent to join up the 'spectator' data with information
from lower energy reactions, both central (one-source) and peripheral
(quasiprojectile). Evidence for a liquid-gas transition has been claimed 
recently for quasiprojectile decay,~\cite{Agostino99}
 as well as spectator decay.~\cite{Pochodzalla95,Elliott98} \

The mechanisms of multifragmentation in an expanding fireball are not yet
fully understood.
The underestimation of the degree of clusterization in a hydodynamic
scenario with {\em local} equilibration, the observed system-size
dependences and the lack of complete isospin-mixing indicate that fast
expansion physics maybe a non-equilibrium process.
While the role of flow seems qualitatively evident, the quantitative
connection to the Equation of State must be further explored.

New isospin-tracing experiments have a potential to overcome some of the
questions relating to a {\em quantitative} assessment of non-equilibrium
features.

The fact that we observe the clusterization of finite pieces of fermionic
matter in at least a partially non-equilibrated environment calls for
application and the further development of {\em quantum} transport
theories, that fulfill in the static limit the well known requirements of
{\em nuclear} physics theory.

\section*{Acknowledgements}
The data and ideas presented here are the result of the coordinated work
of two large Collaborations. 
I owe special thanks to W.~Trautmann for his generous help and advice
concerning the presentation of the ALADIN data.


\end{document}